\documentclass{article} 
\usepackage{iclr2027_conference,times}

\usepackage{amsmath,amsfonts,bm}

\def\eqref#1{equation~\ref{#1}}

\def\1{\bm{1}}

\DeclareMathAlphabet{\mathsfit}{\encodingdefault}{\sfdefault}{m}{sl}
\SetMathAlphabet{\mathsfit}{bold}{\encodingdefault}{\sfdefault}{bx}{n}

\usepackage{hyperref}
\usepackage{url}
\usepackage{booktabs}       
\usepackage{amsfonts}       
\usepackage{nicefrac}       
\usepackage{microtype}      
\usepackage{xcolor}         
\usepackage{graphicx} 
\usepackage{color-edits}
\usepackage{cleveref}
\usepackage{float}
\usepackage[table]{xcolor}
\usepackage{subcaption}
\usepackage{tabularx}
\usepackage{multirow}
\usepackage{amsmath} 
\usepackage{listings}
\usepackage[most]{tcolorbox}
\usepackage[table]{xcolor}
\usepackage{hyperref}
\usepackage{fontawesome5}
\usepackage{graphicx}

\addauthor{vc}{blue}
\addauthor{sh}{red}
\addauthor{ml}{purple}
\addauthor{ph}{orange}

\definecolor{pbframe}{HTML}{8A8A8A}
\definecolor{pbtitle}{HTML}{5A5A5A} 
\definecolor{pbback}{HTML}{F8F8F8}
\definecolor{pbcode}{HTML}{EEEEEF}

\lstdefinestyle{promptcode}{
basicstyle=\ttfamily\scriptsize, breaklines=true, breakatwhitespace=false,
columns=fullflexible, keepspaces=true, showstringspaces=false,
backgroundcolor=\color{pbcode}, frame=none, xleftmargin=2pt, xrightmargin=2pt,
aboveskip=4pt, belowskip=4pt,
} 
\newtcolorbox{promptbox}[1]{%
breakable, enhanced, colback=pbback, colframe=pbframe, boxrule=1pt, arc=2mm,
title={#1}, fonttitle=\bfseries, colbacktitle=pbtitle, coltitle=white,
left=3mm, right=3mm, top=2mm, bottom=2mm, toptitle=1mm, bottomtitle=1mm,
} 

\newcommand{\pbhead}[1]{\par\smallskip\noindent\textbf{#1}\par\nobreak}
\newcommand{\pbsub}[1]{\par\medskip\noindent\textit{\textbf{#1}}\par\nobreak\smallskip}

\newenvironment{pbitemize}{\begin{itemize}\setlength{\itemsep}{0pt}\setlength{\parskip}{0pt}}{\end{itemize}}
\newenvironment{pbenumerate}{\begin{enumerate}\setlength{\itemsep}{0pt}\setlength{\parskip}{0pt}}{\end{enumerate}}

\newcommand{\NumRepos}{35}
\newcommand{\NumInstances}{212}

\newcommand{\AnticipationGap}{foresight resolve rate}

\title{Do Coding Agents Catch Their Own Bugs? (They Don't.)}

\author{Shaswat Patel\textsuperscript{1}, Michel Liao\textsuperscript{2}, Peter Henderson\textsuperscript{2}, He He\textsuperscript{1} \& Valerie Chen\textsuperscript{3} \\
\textsuperscript{1}New York University
\textsuperscript{2}Princeton University
\textsuperscript{3}Stanford University \\
\texttt{spp9399@nyu.edu}
}

\iclrfinalcopy 
\begin{document}

\maketitle

\begin{abstract}
In software development, engineers need to foresee and fix potential problems with their code that are not explicitly specified in requirements---such as pre-emptively considering vulnerabilities or other bugs from a code change.
But current benchmarks often do not measure these future-looking consequences.
We introduce \textbf{Foresight-SWE}, a benchmark for measuring whether coding agents can anticipate and address unstated problems during software engineering tasks.
Foresight-SWE consists of \NumInstances{} samples spanning 4 languages and \NumRepos{} popular open-source repositories.
Each sample contains two pull requests (PRs), the second of which resolves defects the first one introduced. 
To measure whether agents exhibit foresight, we define the \textbf{\AnticipationGap{} (FRR)} as the fraction of instances where an agent fully resolves the original task while also avoiding the defect identified by a later PR.
We evaluate eleven large language models and find that FRR ranges from only 6.13\% to 24.53\%, indicating that models largely fail to address problems introduced by the first PR change.
This raises important questions about how models are trained and aligned in coding contexts, as well as how code review tools should be improved for more fail-safe coding.
Foresight-SWE shifts evaluations in this direction.

\begin{center}
\small
\hypersetup{pdfborder={0 0 0}}

\href{https://github.com/shaswatpatel123/Foresight-SWE}{\faGithub\ \textbf{GitHub}}
\quad\quad
\href{https://huggingface.co/datasets/shaswat123/Foresight-SWE}{%
  \raisebox{-0.2em}{\includegraphics[height=1.3em]{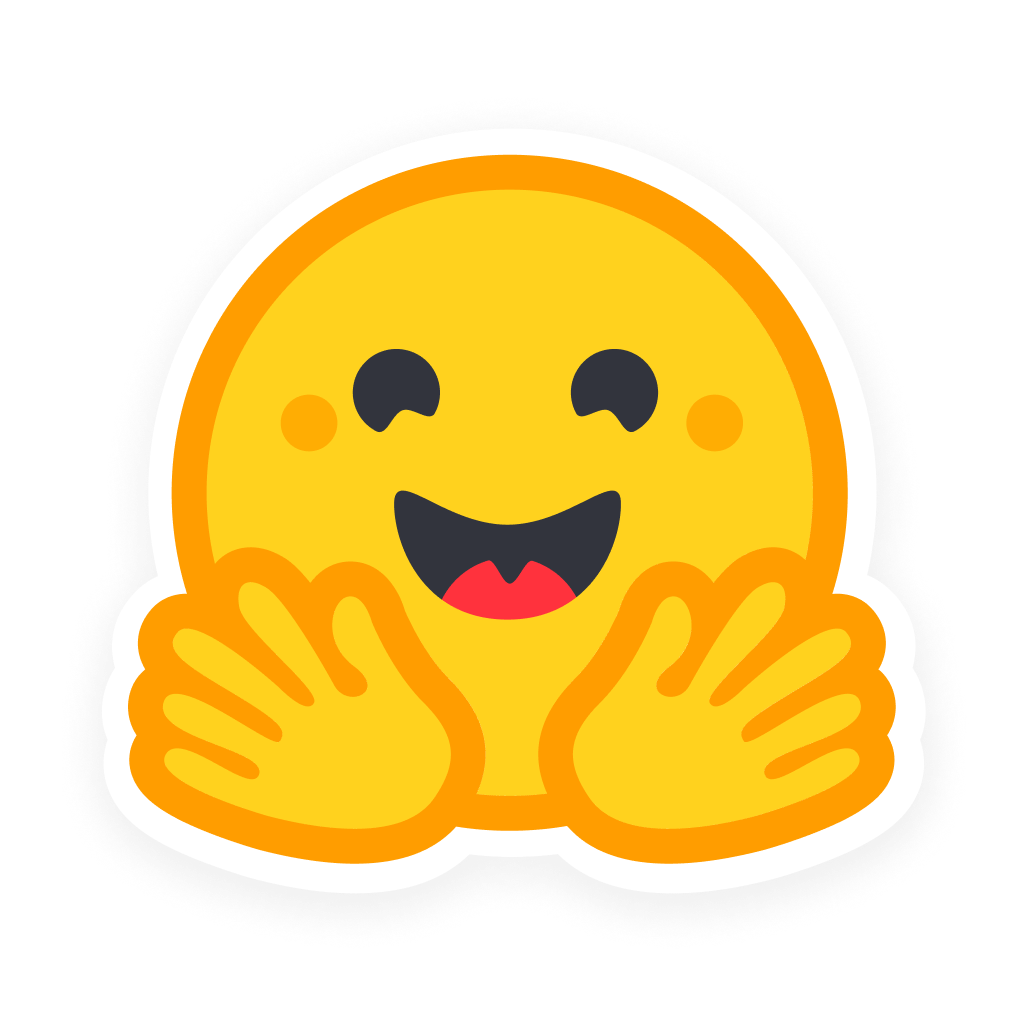}}%
  \textbf{Benchmark}}
\quad\quad
\href{https://shaswatpatel123.github.io/Foresight-SWE/}{\textcolor{yellow!40!orange}{\faTrophy}\ \textbf{Leaderboard}}

\hypersetup{pdfborder={0 0 1}}
\end{center}

\end{abstract}

\section{Introduction}
\begin{figure}[t]
    \centering
    \includegraphics[width=\textwidth]{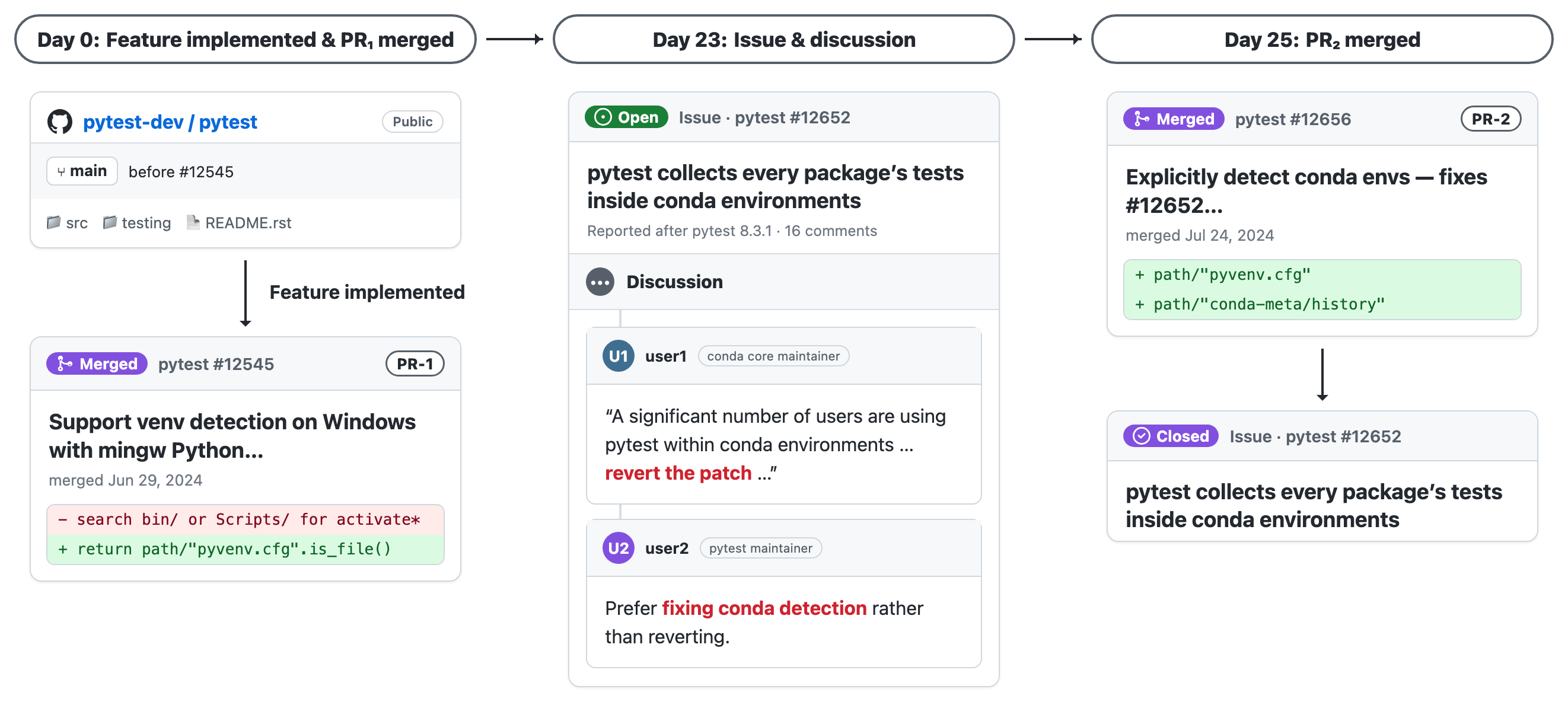}
    \caption{\textbf{Motivating Foresight-SWE.} Consider a real case where, on Day 0, a tested and reviewed $\text{PR}_1$ was merged. By Day 23, users had reported substantial downstream breakage, prompting an urgent discussion that also involved a maintainer of the conda package manager. On Day 25, a follow-up $\text{PR}_2$ repaired the problem and closed the issue. This timeline motivates Foresight-SWE, which measures whether coding agents can foresee consequential failures.}
    \label{fig:concept}
\end{figure}

Coding agents are typically evaluated on software engineering tasks where the complete requirements are explicitly provided \citep{swebench, deng2025swebenchproaiagents, xu2025swecompassunifiedevaluationagentic, swerebenchv2}. Real software tasks, however, are often under-specified \citep{yang2026promptsdontsayunderstanding,zhou2025tom}, and recent work shows that coding-agent performance degrades under such conditions~\citep{larbi2025promptswrongevaluatingcode, laban2025llmslostmultiturnconversation}. As coding agents take on more software engineering work, they should be able to forecast defects that are not explicitly described in the task. Figure~\ref{fig:concept} illustrates such a case. A Pytest contributor changed virtual-environment detection to use Python's standard marker file, passed review, and merged the change. Twenty-three days later, a regression surfaced because conda environments do not create this marker, causing Pytest to incorrectly collect test files from installed packages.

Such latent defects are common and can impose substantial costs on maintainers and users. Django's issue tracker\footnote{\href{https://code.djangoproject.com/}{See here for Django's issue tracker (called Trac).}}
shows 4,225 tickets opened between August 2021 and August 2026, of which 1,988 (47.05\%) were classified as bugs. Of these, 947 were resolved as fixed, amounting to roughly one fixed bug every other day. These bugs involved a median of four participants and five comments, and took a median of 5.3 days from report to fix. For the 22\% not fixed within a month, the median time to resolution rose to 121 days, during which the defect remained live in released code.

Prior work studies related forms of robustness in coding agents, but existing benchmarks either frame the task as proactive debugging in an existing codebase with defects, or restrict evaluation to a narrow class of defects. Active-SWE~\citep{activeswebench} and SWE-Sweep~\citep{swesweep} evaluate whether agents can autonomously identify and repair defects in a repository without issue reports. However, unlike our setting, these benchmarks explicitly instruct agents to search for bugs rather than evaluating whether they proactively anticipate unstated defects to ensure software reliability while implementing a specified task. SecureVibeBench~\citep{securevibebench} evaluates whether agents reproduce vulnerabilities introduced by human developers. However, its evaluation is limited to security vulnerabilities in C/C++ and uses security-specific static and dynamic analyzes, capturing only one aspect of reliable software development. In contrast, we introduce a scalable pipeline that leverages tests from subsequent real-world fixes to evaluate whether coding agents anticipate unstated defects while completing ordinary software development tasks, spanning diverse defect types and programming languages.

We introduce \textbf{Foresight-SWE} to measure whether coding agents can anticipate and address unstated edge cases while implementing a scoped task.  
To construct this benchmark, we mine PRs from open-source repositories and look for candidate pairs where one PR makes a change and a later PR fixes an edge case it missed. 
To identify valid pairs, we combine explicit maintainer references with information based on code history to reliably trace changes back to the earlier PRs. 
We then reconstruct executable repository states for both PRs, and validate each pair through test execution. 
We start from 11,609 candidate PR pairs, of which 492 remain after our construction and validation pipeline. From these validated pairs we introduce Foresight-SWE, which consists of \NumInstances{} instances spanning four programming languages and \NumRepos{} popular open-source repositories.

We find all tested models struggle with foresight compared to resolving the original task.
On average, agents resolve the original task in 60.7\% of instances.
However, when considering \textbf{\AnticipationGap{} (FRR)}---the fraction of instances where the agent both resolves the original task and anticipates the defect later repaired by the follow-up PR---models achieve an average FRR of only 14.8\%. When models are told the defect specifications,  
models can achieve a much higher average resolve rate of 91.2\%, indicating the challenge is in recognizing the unstated defect. We find that proactive prompting can improve foresight (e.g., FRR increases by 11.32 percentage points for Claude Opus 5) but often trades off with the resolve rate of the original task.
Surprisingly, in contrast, popular code review skills are not able to increase FRR.
These results suggest that current agents are often capable of fixing a defect once it is known, but are much less reliable at recognizing during implementation that their own change introduces a failure mode that will surface later. 

\textbf{Contributions.} First, we develop a scalable pipeline that mines real-world repositories, inspects their code history to produce paired instances of an original software change and a later fix for a defect introduced by that change. Second,  we introduce Foresight-SWE, a benchmark built from real PR pairs where an initial change introduces a defect that is repaired only by a later PR.  Third, we show across eleven models that they are much worse at foresight than task completion and realistic prompt interventions are not necessarily a reliable way of counteracting lack of foresight. 

\section{Related Work}\label{sec:related_work}
\textbf{Agents evaluated on multi-PR evolution.} Benchmarks such as SWE-bench~\citep{swebench} and its successors~\citep{deng2025swebenchproaiagents, xu2025swecompassunifiedevaluationagentic} evaluate agents on isolated issues, while recent work extends evaluation to sequences of dependent changes, examining both task completion~\citep{swemilestone, swechain, chainswe, sweci} and long-term code quality and maintainability~\citep{slopcodebench, codethread}. Recent work has also leveraged PR dependencies to train long-horizon agents~\citep{davinciagency,weave}. These approaches typically focus on implementing explicitly specified changes while preserving correctness across successive modifications. In contrast, our work requires agents to proactively identify and address future defects that are not specified in the original task. Furthermore, rather than executing agents across a sequence of PRs, our setup evaluates them only on the initial task and uses tests from future PRs to assess whether their initial implementation anticipates future defects.

\textbf{Agents deployed to find bugs.} Recent works have focused on building an agent to search and fix defects. Active-SWE~\citep{activeswebench} and SWE-Sweep~\citep{swesweep} withholds the issue report and asks agents to find and fix bugs from a repository snapshot, reporting a best resolved rate of 20.0\% and 9.5\%, respectively. RepoAudit~\citep{repoaudit} isolates identification by never repairing, reporting 40 true positives at 78.43\% precision. Live-kBench~\citep{livekbench} supplies a crash report and reproducer and still reports a low equivalent-patch rate against the developer fix, with partial fix and latent bug surfacing as named failure modes.
SWE-Review~\citep{swereviewbench} evaluates residual-defect detection once a candidate patch exists, and the security-lifecycle line~\citep{cybergym, cvebench} presupposes a known vulnerability and measures PoC construction or repair. In all of these prior works, the agent knows it is looking for bugs. 
We expect coding agents to actively forecast potential defects introduced by their changes as part of the implementation process, without necessarily deploying a separate approach as in prior work.

\textbf{Defect prediction before implementation.} Another relevant line of work predicts where defects will appear without generating code. Just-in-time Software Defect Prediction (JIT-SDP) estimates whether a code change is likely to introduce a defect using properties of the code change and its development history, such as its size, the files it touches, developer experience~\citep{kamei,zhao}. Ticket-level prediction moves the defect prediction point earlier in the development lifecycle, estimating whether implementing a software ticket will introduce a defect before any code is written~\citep{ticketlevelbug}. ReDef extends JIT-SDP to LLM models, using them to classify individual code changes as defect-inducing or clean~\citep{redef}, while CodeFlowLM further studies which changed lines are responsible for the defect~\citep{codeflowlm}. Our work goes beyond predicting defect-prone changes by measuring whether an agent can identify an unstated failure mode and prevent the resulting defect while implementing the task itself.

\textbf{Forecasting future outcomes.} Forecasting has been studied broadly as a way to evaluate whether LLMs can reason about future outcomes from currently available evidence. 
For example, ForecastBench~\citep{forecastbench} and MIRAI \citep{mirai} evaluate forecasting of future events from historical information. 
Fewer works have considered forecasting in software engineering use cases. 
SecureVibeBench~\citep{securevibebench} measures whether agents reproduce the same security vulnerability introduced by the original developer for a limited set of C/C++ projects, capturing only one aspect of reliable software development. 
In contrast, we develop a general-purpose pipeline to mine PRs where coding agents need to foresee and fix potential problems with their code, capturing a broader range of software defects, including silent incorrect behavior, across multiple programming languages.

\section{Foresight-SWE}\label{sec:foresightbench}

\begin{figure}[t]
    \centering
    \includegraphics[width=\textwidth]{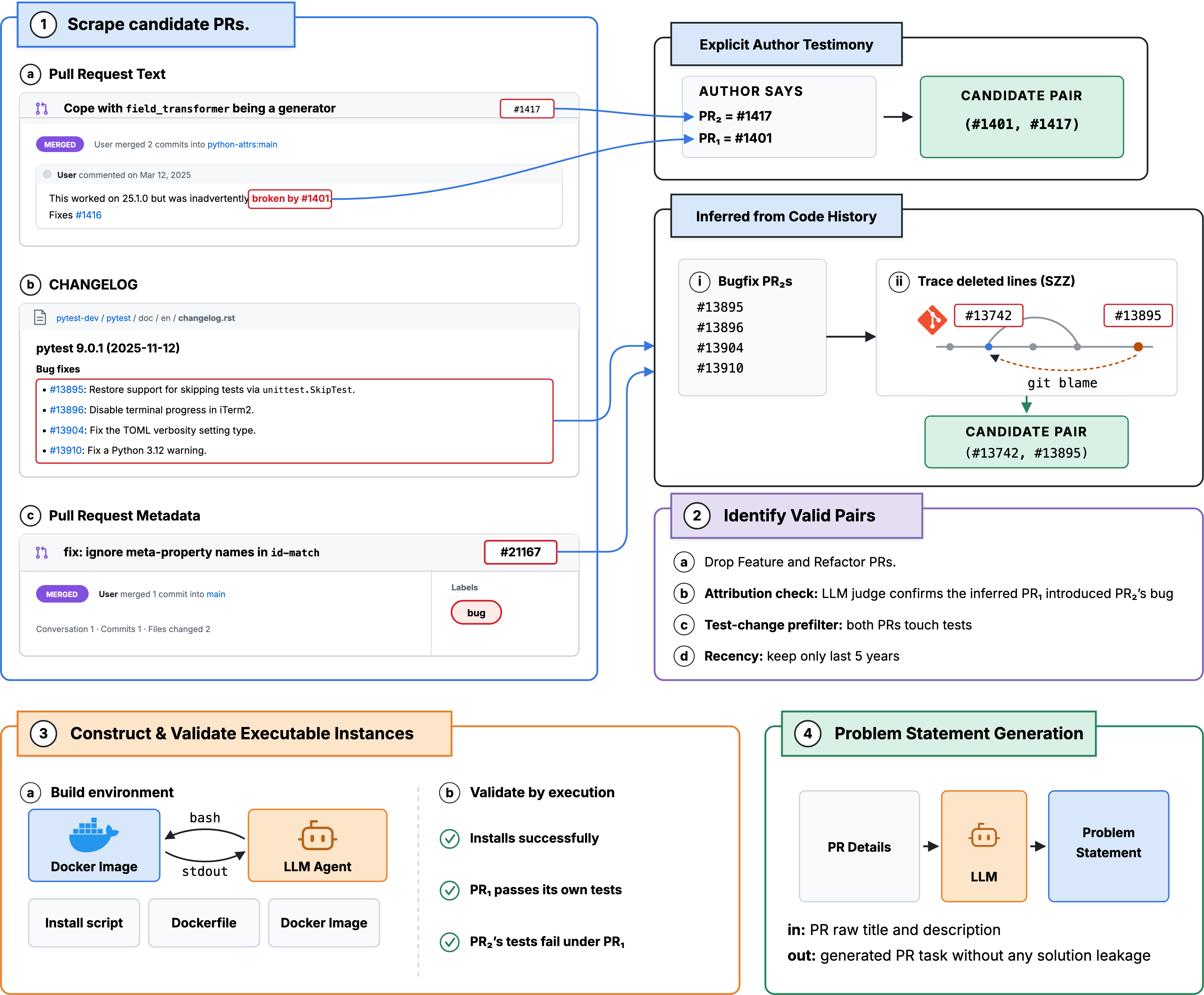}
    \caption{\textbf{Overview of Foresight-SWE Construction.} 
    We scrape candidate ($\text{PR}_1$, $\text{PR}_2$) pairs from GitHub through either explicit author testimony or by inferring from code history. The candidates then pass through filtering to identify valid pairs, environment synthesis, and finally execution filtering.}
    \label{fig:dataset_construction_pipeline}
\end{figure}

We outline the process of constructing Foresight-SWE, as illustrated in Figure~\ref{fig:dataset_construction_pipeline}. 
The pipeline identifies software changes where an original PR successfully addresses its stated task but leaves behind a defect that is repaired by a later PR.
See Appendix \ref{appendix:benchmark_construction} for more details on benchmark construction and validation.

\subsection{Benchmark construction}

\textbf{Step 1: Scrape candidate PRs.} 
We identify pairs of PRs from the same repository such that $\text{PR}_2$ repairs a defect introduced by $\text{PR}_1$ using two approaches. First, we use explicit author testimony, such as ``introduced in $\text{PR}_1$," ``broken by $\text{PR}_1$," or ``follow-up to $\text{PR}_1$." 
Explicit references to the origin of a defect depend heavily on repository conventions and maintainer practices. 
Since many repositories do not contain explicit references, we also identify candidate pairs using changelogs, Trac tickets (for Django), and CVE/GHSA advisories, and subsequently use SZZ~\citep{sliwerski2005szz} to infer $\text{PR}_1$. 
SZZ traces the lines deleted or rewritten by $\text{PR}_2$ back to the commits that introduced them. 
It then maps these commits to PRs and selects the PR responsible for introducing the largest number of affected lines as the candidate $\text{PR}_1$. 
While this process isn't guaranteed to be exhaustive, it captures many bugs that were critical enough to be raised and resolved by humans in real-world codebases. The initial pool identified by this process had 11,609 candidates across 35 repositories and four programming languages.

\textbf{Step 2: Identify valid foresight pairs.} From the set of mined pairs in Step 1, we want to ensure that they satisfy the following criteria: First, $\text{PR}_2$ must repair a genuine defect caused by $\text{PR}_1$, and that defect must be a reasonable target for anticipation. We therefore exclude cases where $\text{PR}_2$ primarily implements a new feature, performs a refactor, simply reverts $\text{PR}_1$, or entangles the defect fix with substantial unrelated changes. Second, when $\text{PR}_1$ is inferred using SZZ, the inferred PR must be the change that actually caused the defect. We apply two LLM-as-a-judge filters to enforce these criteria. We first judge whether $\text{PR}_2$ repairs an anticipatable defect associated with $\text{PR}_1$~\citep{swerebenchv2}. 
We then judge only SZZ-inferred pairs and verify that the inferred $\text{PR}_1$ is indeed the PR responsible for introducing the defect. 
Importantly, we require both PRs to modify test files so that $\text{PR}_1$ can be scored on the original task and $\text{PR}_2$ can provide anticipation tests.
Finally, we restrict both PRs to a five-year window from 2021-2026 to ensure repositories can be reliably rebuilt. After filtering, 2,262 pairs remain. 

\textbf{Step 3: Construct and validate executable instances.} Evaluating each PR pair requires reproducing both the codebase and its execution environment, so we build isolated containerized environments for all selected PRs. For each instance, we reconstruct the repository at four states: the base and head of $\text{PR}_1$, and the base and head of $\text{PR}_2$. A PR's base is the repository state immediately before the PR is applied, while its head is the state after the PR's changes are applied. Following SWE-ReBench~V2~\citep{swerebenchv2}, we use an LLM agent to infer the installation and test commands needed to execute these states. We then replay each generated environment from scratch and validate the PR pair by executing its tests across the reconstructed states. $\text{PR}_1$ and $\text{PR}_2$'s tests must fail at its base and pass at its head. The $\text{PR}_2$'s test must also fail at $\text{PR}_1$ head, confirming that the defect was already present after $\text{PR}_1$. We retain only instances which satisfy these conditions, and produce machine-readable outputs, leaving 815 pairs. 

\textbf{Step 4: Problem Statement Generation.} Raw PR titles and descriptions are not always suitable as standalone benchmark tasks because they may be incomplete, rely on discussion context, or expose implementation details that bias an agent toward the developer’s solution~\citep{swebench, swerebenchv2, xu2025swecompassunifiedevaluationagentic}. Following recent benchmarks~\citep{swerebenchv2, xu2025swecompassunifiedevaluationagentic, swechain}, we generate a self-contained problem statement for $\text{PR}_1$ using an LLM conditioned on its title, description, and source patch. 
We verify that the new problem statement (1) describes the intended behavior without prescribing the reference implementation or revealing the latent defect, and (2) is sufficiently specified for an agent to satisfy $\text{PR}_1$'s tests. 
We discard instances that fail either criterion. After validation, 492 instances remain, from which we draw a stratified evaluation set of 212. 

\begin{figure*}[t]
    \centering

    \begin{minipage}[t]{0.35\textwidth}
    \vspace{0pt}
    \centering\footnotesize
    \setlength{\tabcolsep}{3.5pt}
    \begin{tabular}{@{}llrrr@{}}
        \toprule
             & Attribute & Mean & Med. & Max \\
        \midrule
        \multirow{5}{*}{$\text{PR}_1$}
          & Lines edited & 367 &    99 &  10833 \\
          & Files edited  &   5.8 &     3 &     86 \\
          & Hunks        &  18.4 &     8 &    245 \\
          & F2P tests    &  20.3 &     3 &    666 \\
          & P2P tests   &  1423 &   151 &  28074 \\
        \midrule
        \multirow{4}{*}{$\text{PR}_2$}
          & Lines edited &  49.8 &    15 &   1403 \\
          & Files edited &   1.8 &     1 &     41 \\
          & Hunks        &   5.2 &     2 &    191 \\
          & F2P tests &   4.5 &     2 &    194 \\
        \midrule
        \multicolumn{2}{@{}l}{Surface time (days)} & 158 & 88 & 1215 \\
        \bottomrule
    \end{tabular}

    \end{minipage}
    \hfill
    \begin{minipage}[t]{0.605\textwidth}
        \vspace{0pt}
        \centering
        \includegraphics[width=\linewidth]{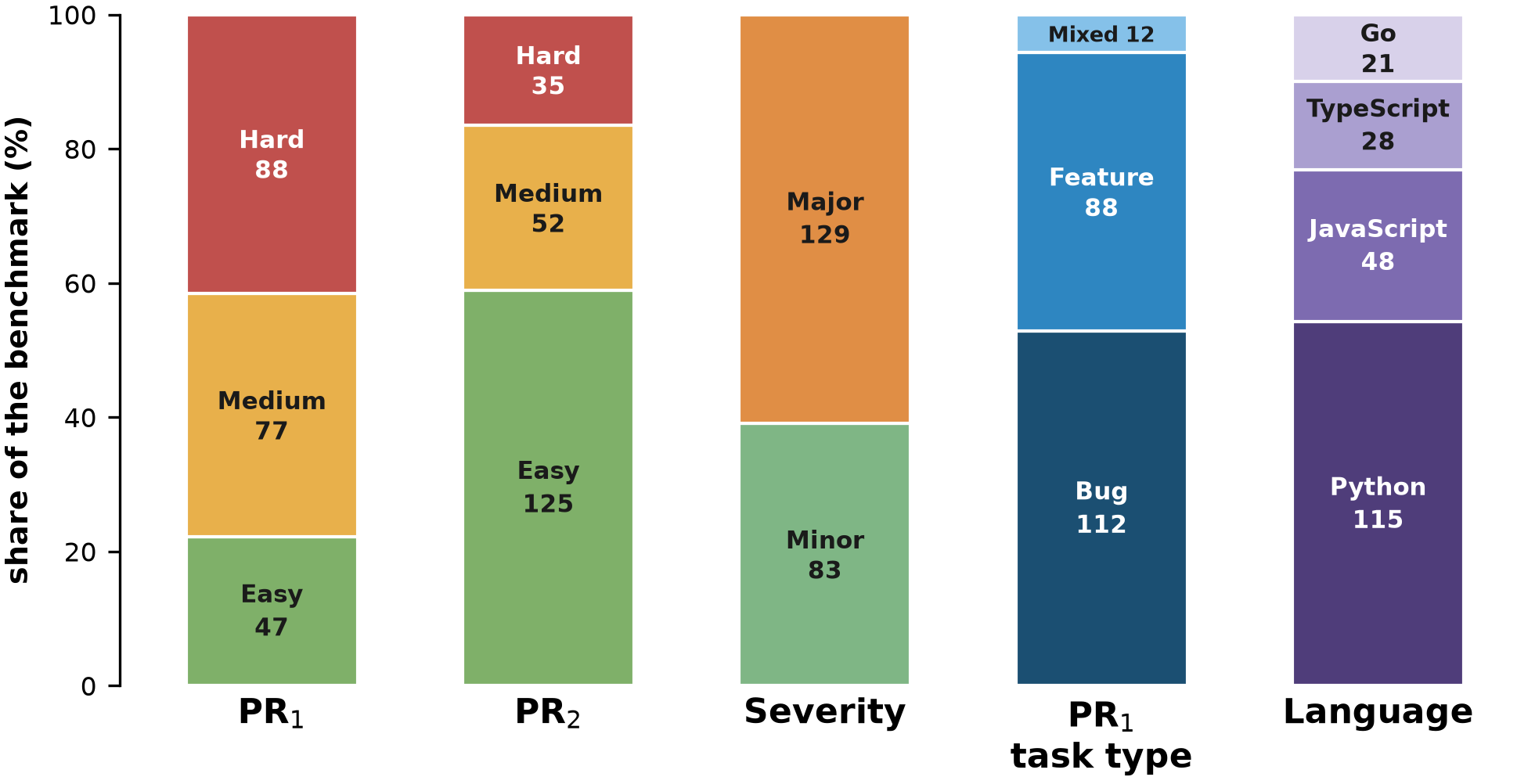}
    \end{minipage}

    \caption{\textbf{Foresight-SWE dataset distribution.} The benchmark contains \NumInstances{} instances from \NumRepos{} repositories across four languages. (Left) Summary statistics of both $\text{PR}_1$ and $\text{PR}_2$. (Right) Distribution of problem difficulty, severity, task type, and programming language.
    }
    \label{fig:dataset-distribution}
\end{figure*}

\subsection{Dataset Overview} 

Foresight-SWE currently comprises \NumInstances{} instances from \NumRepos{} repositories across four languages, containing 958 anticipation tests, 4,307 task tests, and 301,659 regression tests (Figure~\ref{fig:dataset-distribution}). 
We observe that a median of 88 days separates $\text{PR}_1$ from $\text{PR}_2$ (mean 158, maximum 1,215), so these are defects that survived review and release.

\textbf{Task difficulty.} We annotate each instance with task difficulty and defect severity following the approach in SWE-ReBench~V2~\citep{swerebenchv2}. Difficulty reflects the expected implementation effort based on the amount of code change, logic complexity, and required domain knowledge. We classify tasks as easy if they are expected to take less than 15 minutes, medium if they require 15 minutes to 1 hour, and hard if they require more than 1 hour. The $\text{PR}_1$ tasks skew harder, with 47 easy, 77 medium, and 88 hard instances, while $\text{PR}_2$ skews easier, with 125 easy, 52 medium, and 35 hard instances. This indicates that the missed defects are often substantially easier to repair than the original tasks were to implement. 

\textbf{Defect severity.} We classify severity into three categories, adapting common software severity schemes that consider user impact and the availability of workarounds~\citep{arethesebugsreallynormal, detectSilentFailures}. Critical defects cause severe failures on common paths, such as crashes, data corruption, or security bypasses. Major defects silently produce incorrect behavior for majority of users. Minor defects are explicit, narrowly triggered, low reach, easily worked around, or cosmetic. We use LLM-as-a-judge to annotate each instance. Among the \NumInstances{} instances, 124 are classified as major, 83 as minor, and 5 as critical.
And, in 129 of the 212 instances, the $\text{PR}_2$ test fails on an assertion rather than an exception, indicating a silent incorrect result rather than an explicit failure. 

We provide the prompts used for annotation of task difficulty and defect severity in Appendix~\ref{appendix:data_annotation}.

\section{Results}\label{sec:results_sec}

We first describe the experimental set-up and then present the results, which comprise evaluations of 11 models on Foresight-SWE and subsequent experiments to explain where the gap in implementation and foresight arises from.

\subsection{Experimental Set up}

\textbf{Evaluation protocol.} The agent receives the repository at $\text{PR}_1$ together with the generated task description, with no information about $\text{PR}_2$ or the later defect. The agent produces a patch, which we evaluate on tests for the original task and on held-out tests for the defect later repaired by $\text{PR}_2$.

\textbf{Metrics.} Following \citet{swebench}, we define the resolve rate as the fraction of instances for which an agent passes all $\text{PR}_1$ tests. These tests are divided into two categories: Fail-to-Pass (F2P) tests, which fail before $\text{PR}_1$ and pass after it, and Pass-to-Pass (P2P) tests, which already pass before $\text{PR}_1$ and must continue to pass after the agent's change. Together, they measure whether the agent implements the requested behavior without introducing regressions. To measure foresight, we define the \textbf{anticipation score} as the mean fraction of $\text{PR}_2$ F2P tests passed. These tests capture the defect later repaired by $\text{PR}_2$, so higher scores indicate that the agent handled more of the unstated behavior associated with that defect. Finally, we define our primary metric, which is the \textbf{foresight resolve rate (FRR)}: the fraction of instances for which the agent passes all $\text{PR}_1$ F2P and P2P tests and all $\text{PR}_2$ F2P tests. FRR captures the desired behavior, measuring \textit{both} how often an agent both resolves the original task and avoids the later defect. 

\textbf{Models.} We evaluate 11 models, including 6 open-weight and 5 closed models. For closed-weight models, we use the providers' native coding-agent harnesses, such as Codex~\citep{codex} for OpenAI models and Claude Code (CC)~\citep{claudecode} for Anthropic models, since these are the interfaces through which users interact mainly and are optimized for their respective models. Open-weight models generally do not have an equivalent standardized native harness, so we evaluate them under a common scaffold using mini-swe-agent (MSA)~\citep{yang2024sweagent}, a lightweight harness widely used in coding-agent evaluation~\citep{programbench, activeswebench, xu2025swecompassunifiedevaluationagentic}. We provide additional budget and version details in Appendix~\ref{appendix:exp_setup_detail}.

\subsection{Foresight Performance Across Models}\label{sec:main_results}

\begin{table}[t]
    \centering
    \caption{\textbf{Agents rarely anticipate later defects.}
    Across eleven models, agents achieve an average resolve rate of 60.68\%, but an average FRR of only 14.84\%. Their average anticipation score is 26.39\%, indicating that agents often capture only part of the behavior required to avoid the later defect. Bold represents the best-scoring model for each column's metric.
    }
    \begin{tabular}{llccc}
      \toprule
      Model & Resolve Rate ($\uparrow$) & Ant. Score ($\uparrow$) & \textbf{FRR ($\uparrow$)} 
      \\
      \midrule
      GLM-5.3 Flash          & 80.66 & 19.64 & 11.32 \\
      Qwen3.8 Flash Next     & \textbf{91.98} & 22.88 & 20.75 \\
      Qwen3.5-397B-A17B      & 21.70 & 26.44 &  7.08 \\
      DeepSeek-V4 Flash      & 58.49 & \textbf{31.76} & 18.87 \\
      Qwen3.8-27B            & 25.47 & 28.37 &  7.55 \\
      Qwen3.5-27B            & 18.40 & 26.00 &  6.13 \\
      \midrule
      GPT-5.6 Sol            & 84.43 & 23.98 & 18.40 \\
      GPT-5.6 Terra          & 74.53 & 23.91 & 15.09 \\
      GPT-5.6 Luna           & 74.06 & 28.01 & 16.51 \\
      \midrule
      Claude Sonnet 5        & 59.91 & 28.76 & 16.98 \\
      Claude Opus 5          & 77.83 & 30.57 & \textbf{24.53} \\
      \bottomrule
    \end{tabular}
    \label{tab:main_results}
\end{table}

\begin{figure}
    \centering
    \includegraphics[width=\linewidth]{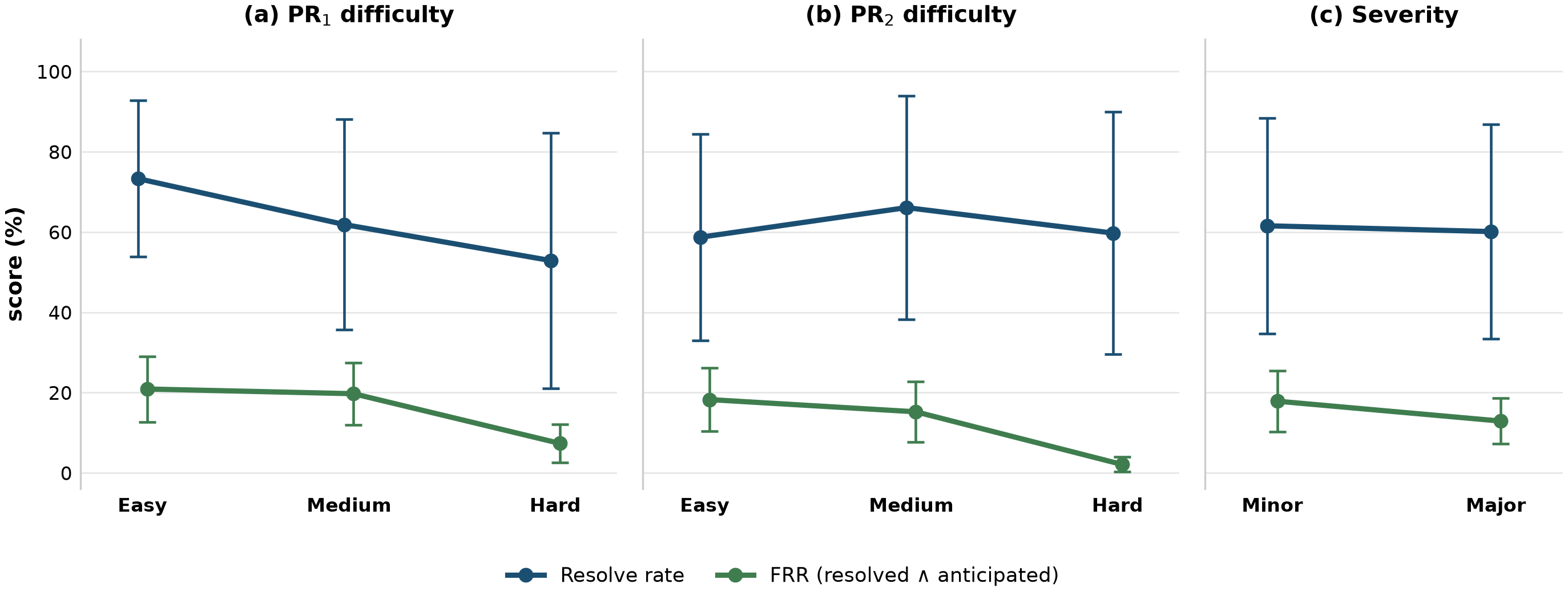}
    \caption{\textbf{Foresight decreases as task difficulty and defect severity increase.} We stratify the \NumInstances{} instances along three axes. Across $\text{PR}_2$ difficulty (b) FRR degrade dramatically but resolve rate stays mostly stable. Across $\text{PR}_1$ difficulty (a), all metrics degrade. Defect severity (c) has little effect on resolve rate, but FRR is 5.0 percentage points lower for major defects, with nine of eleven models performing better on minor defects. Error bars show $\pm 1$ SD across models.}
    \label{fig:strati_results_v1}
\end{figure}

\textbf{Agents resolve the stated task far more often than they anticipate the later defect.}\footnote{
Data contamination is not an issue for Foresight-SWE. If a model has memorized $\text{PR}_2$, it would know of the $\text{PR}_2$ test it is being evaluated on, thereby inflating its FRR. However, the best-performing frontier model achieves an FRR of only 24.53\%. If a model has memorized $\text{PR}_1$, then foresight becomes a larger issue since the model is reintroducing a known defect.
} Across the 11 models, the average resolve rate is 60.7\%, while the average FRR is only 14.8\% (Table~\ref{tab:main_results}). Even among stronger task-solving agents, foresight is much weaker. For example, Qwen3.8 Flash Next achieves the highest resolve rate at 92.0\%, but an FRR of only 20.8\%. Claude Opus 5 attains the highest FRR at 24.5\%, meaning that no model both resolves the original task and avoids the later defect on more than a quarter of instances. 
Recall that 59\% of PR$_2$ tasks are easy (\Cref{fig:dataset-distribution}), and yet we still see low FRR.
In one instance, an agent is asked to change the order of operations when constructing a class from its declared fields. 
The library originally validated the field ordering before running a user-supplied callback function, and users frequently supplied a generator for the callback function. 
In Python, a generator output can only be read once and a second attempt to read it yields no values. However, reordering the operations causes the generator output to be consumed twice. 
While the $\text{PR}_2$ fix is a single-line change that wraps the generator output in a tuple, all eleven models failed this instance.

\textbf{Foresight degrades with difficulty and severity.}
Stratifying results by difficulty (Figure~\ref{fig:strati_results_v1}) reveals the difference between task completion capabilities and defect anticipation. Across all three $\text{PR}_2$ difficulty, resolve rate across all models does not decrease monotonically. In contrast, FRR falls from 18.3\% to 2.1\%, all models score lower on hard defects than on easy ones, and, in particular, GPT 5.6 Sol scored zero on the hard difficulty defects.
Across $\text{PR}_1$ difficulty, both task completion and foresight decline.  
We also observe average degradation in FRR of 5.0\% when comparing minor and major defects, while the resolve rate shows no degradation. As part of annotating instances for severity, we also label instances as silent (the code runs and returns a wrong answer without any error trace) or loud (the code raises an error). Prior work has illustrated that silent failures often require deeper reasoning about intended behavior and greater codebase knowledge to identify~\citep{detectSilentFailures, demystifyingSilentErrors}. We find that agents achieve an FRR of 18.5\% on loud defects, compared with only 12.5\% on silent defects, suggesting that defects without an explicit failure signal are substantially harder to anticipate. We also observe variation in foresight performance across programming languages (Figure~\ref{fig:lang_strat_results}). Per-model stratified results for difficulty, severity, and detectability are provided in Appendix~\ref{appendix:strati_full_result}.

\subsection{Is the gap identification or implementation?}\label{sec:capfore}

\begin{table}[t]
    \centering
    \caption{\textbf{Foresight is an issue of identification.} When the defect is explicitly specified, agents receive $\text{PR}_2$'s task description and resolve 91.23\% of instances on average. When the same defects must instead be anticipated from the original task, the models achieve an average FRR of only 19.91\%. We conduct this ablation on a subset of 5 open-weight and closed models.}
    \begin{tabular}{lccc}
      \toprule
      & \multicolumn{1}{c}{Defect specified}
      & \multicolumn{2}{c}{Defect unspecified} \\
      \cmidrule(lr){2-2}\cmidrule(lr){3-4}
      Model & Resolve Rate ($\uparrow$) & Ant. Score ($\uparrow$) & \textbf{FRR ($\uparrow$)} \\
      \midrule
      Qwen3.8 Flash       & 97.64 & 22.88 & 20.75 \\
      DeepSeekV4 Flash    & 85.85 & 31.76 & 18.87 \\
      GPT-5.6 Sol         & 94.81 & 23.98 & 18.40 \\
      Claude Sonnet 5     & 86.32 & 28.76 & 16.98 \\
      Claude Opus 5       & 91.51 & 30.57 & 24.53 \\
      \bottomrule
    \end{tabular}
    \label{tab:pr2direct}
\end{table}

A low foresight score may either be caused by the agent being unable to implement the fix or failing to foresee and identify the defect. 

\textbf{Setup.} To separate defect identification from defect resolution, we evaluate five models on the $\text{PR}_2$ task. The agent receives the $\text{PR}_2$ task description and starts from $\text{PR}_2$'s base commit, measuring how well it can repair the defect when the defect is explicitly specified. We compare this performance with the original foresight setting, where the agent receives only the $\text{PR}_1$ task and must both identify and avoid the defect, as measured by FRR and Foresight Score. For this experiment, we select the two open-weight models with the highest FRR in the main evaluation together with three frontier closed-weight models.

\textbf{Results.} Table~\ref{tab:pr2direct} shows a large gap between fixing a known defect and anticipating it from the original task. When the defect is explicitly specified, models resolve 91.23\% of instances on average. In the foresight setting, where the defect is not revealed, the same models achieve an average FRR of only 19.91\% and a Foresight Score of 27.59\%. This pattern holds across all five models, despite substantial differences in their ability to solve the original task. 
These results suggest that the primary bottleneck is identifying the latent defect rather than implementing its repair once identified.

\subsection{Can prompting alone close the gap?}\label{sec:prompting}

Section~\ref{sec:capfore} shows that models can usually repair a defect once it is explicitly described. We therefore ask whether simple and realistic prompting-based interventions can elicit more foresight.

  \begin{table}[t]
      \centering
      \caption{\textbf{Prompting improves foresight, but reduces task completion.} For a subset of models, we compare the baseline against proactive prompting, self-review, and their combination. The FRR improves, but the resolve rate drops steadily across different techniques. 
      Bold represents the best setting for each model and column.
      }
      \begin{tabular}{llccc}
        \toprule
        Model & Setting & Resolve Rate ($\uparrow$) & Ant. Score ($\uparrow$) & \textbf{FRR ($\uparrow$)} \\
        \midrule
        \multirow{4}{*}{GPT-5.6 Sol}
          & Baseline                   & \textbf{84.43} & 23.98 & 18.40 \\
          & + Proactive prompt         & 82.55 & 39.77 & 31.60 \\
          & + Self-review              & 69.81 & 34.10 & 24.06 \\
          & + Proactive \& self-review & 70.28 & \textbf{44.90} & \textbf{32.55} \\
        \midrule
        \multirow{4}{*}{Claude Opus 5}
          & Baseline                   & \textbf{77.83} & 30.57 & 24.53 \\
          & + Proactive prompt         & 75.47 & 44.00 & 32.55 \\
          & + Self-review              & 68.40 & 46.38 & \textbf{35.85} \\
          & + Proactive \& self-review & 64.62 & \textbf{52.97} & \textbf{35.85} \\
        \midrule
        \multirow{4}{*}{Claude Sonnet 5} 
          & Baseline                   & \textbf{59.91} & 28.76 & 16.98 \\ 
          & + Proactive prompt         & 44.81 & 31.87 & 13.68 \\
          & + Self-review              & 57.08 & 30.47 & \textbf{18.40} \\
          & + Proactive \& self-review & 42.45 & \textbf{32.81} & 14.15 \\
        \bottomrule
      \end{tabular}
      \label{tab:prompt_variants}      
  \end{table} 

\textbf{Setup.} We test three interventions that target different stages of the agent's workflow. The first, a proactive prompt, modifies the initial task prompt by explicitly asking the agent to consider edge cases and missing details that may not be stated in the problem description. The second, a self-review prompt, adds a self-review step after the agent submits its patch. We invoke the harness's native review command, \texttt{/self-review} for Claude Code and \texttt{/review} for Codex, and ask the same agent to inspect its completed patch and repair any additional issues it identifies. The proactive prompt tests whether agents fail to search beyond the stated requirements during implementation, while self-review tests whether they stop once the explicit task appears complete. The third 
combines both approaches, applying the proactive prompt during implementation followed by self-review after the patch is completed.

\textbf{Results.} Table~\ref{tab:prompt_variants} shows that both interventions can recover part of the foresight gap, but often at the cost of task completion. Proactive prompting raises FRR for GPT-5.6 Sol and Claude Opus 5 by 13.20 and 8.02 points, respectively, while reducing resolve rate by only 1.88 and 2.36 points. However, even the best-performing model, Claude Opus 5, reaches an FRR of only 35.85\%, far below its 91.51\% resolve rate when the defect is explicitly specified. Thus, prompting recovers only part of the foresight that the model is capable of once the defect is known. In contrast, the native review mechanisms in Claude Code and Codex increase FRR for all three models but reduce resolve rate in every case, and in case of GPT-5.6 Sol it decreases by 14.62 points. This suggests that broader review can uncover unstated defects while also introducing regressions in the original task.  
A qualitative example from \texttt{aiohttp} illustrates this behavior. $\text{PR}_1$ fixes an HTTPS defect where canceled requests could release a connection too early, but in doing so routes \texttt{close()} through \texttt{release()} logic. $\text{PR}_2$ later restores the distinction between releasing and closing a connection. 
Baseline Claude Opus 5 reproduces $\text{PR}_1$'s approach, resolving $\text{PR}_1$ but missing the defect.  With proactive prompting, Opus anticipates the $\text{PR}_2$ fix, but makes additional cleanup changes that cause $\text{PR}_1$ tests to fail. 

\section{Conclusion}
As software development increasingly adopts coding agents, it is important to measure whether they can foresee defects that are unstated in the task. We develop a dataset construction pipeline that collects real-world GitHub pull requests spanning four programming languages and 35 popular open-source repositories. Across 11 models, agents achieve an average resolve rate of 60.7\%, but an average foresight resolve rate of only 14.8\%, showing that anticipating unstated defects remains challenging even for current state-of-the-art models. Our findings highlight the need to train future coding agents not only to satisfy stated requirements, but also to anticipate potential failure modes and produce more robust implementations.

\textbf{Limitations and Future Work.}
Foresight-SWE includes only defects that were eventually identified by maintainers, fixed, and successfully reproduced through execution. While this setup is sufficient for studying whether agents can anticipate unstated defects in the task description, it excludes defects that remain undiscovered by human maintainers. Our FRR metric rewards agents for handling latent defects but does not penalize unnecessary defensive changes. An agent that anticipates many defects may still produce low-quality or overly generalized code. Future work could therefore develop more robust evaluation metrics that jointly capture defect prevention and code quality, while penalizing excessive generalization. Finally, the benchmark can be expanded to cover additional languages and defect types.

\section{AI use statement}
We used AI-based tools (Claude Code and Codex) to generate code to populate plots and tables. We also used AI tools for grammar checking and minor language editing. 

\section{Ethics statement}
This work introduces Foresight-SWE, a benchmark for evaluating whether coding agents can anticipate and prevent latent software defects during implementation. We state that this work is intended for ethical and constructive purposes. While no immediate societal harms are identified, coding agents should be evaluated and deployed with adequate oversight to mitigate potential misuse and unintended software failures.

\section{Reproducibility statement}
We describe the dataset construction process in Section~\ref{sec:foresightbench} and the experimental settings in Section~\ref{sec:results_sec}. More details including prompts, and experimental setting are provided in the Appendix. We will also make our benchmark and codebase public in the future.

\bibliography{iclr2027_conference}

@misc{xu2025swecompassunifiedevaluationagentic,
      title={SWE-Compass: Towards Unified Evaluation of Agentic Coding Abilities for Large Language Models}, 
      author={Jingxuan Xu and Ken Deng and Weihao Li and Songwei Yu and Huaixi Tang and Haoyang Huang and Zhiyi Lai and Zizheng Zhan and Yanan Wu and Chenchen Zhang and Kepeng Lei and Yifan Yao and Xinping Lei and Wenqiang Zhu and Zongxian Feng and Han Li and Junqi Xiong and Dailin Li and Zuchen Gao and Kun Wu and Wen Xiang and Ziqi Zhan and Yuanxing Zhang and Wuxuan Gong and Ziyuan Gao and Guanxiang Wang and Yirong Xue and Mengtong Li and Mengfei Xie and Xiaojiang Zhang and Jinghui Wang and Wenhao Zhuang and Zheng Lin and Huiming Wang and Zhaoxiang Zhang and Yuqun Zhang and Haotian Zhang and Bin Chen and Jiaheng Liu},
      year={2025},
      eprint={2511.05459},
      archivePrefix={arXiv},
      primaryClass={cs.SE},
      url={https://arxiv.org/abs/2511.05459}, 
}

@misc{deng2025swebenchproaiagents,
      title={SWE-Bench Pro: Can AI Agents Solve Long-Horizon Software Engineering Tasks?}, 
      author={Xiang Deng and Jeff Da and Edwin Pan and Yannis Yiming He and Charles Ide and Kanak Garg and Niklas Lauffer and Andrew Park and Nitin Pasari and Chetan Rane and Karmini Sampath and Maya Krishnan and Srivatsa Kundurthy and Sean Hendryx and Zifan Wang and Vijay Bharadwaj and Jeff Holm and Raja Aluri and Chen Bo Calvin Zhang and Noah Jacobson and Bing Liu and Brad Kenstler},
      year={2025},
      eprint={2509.16941},
      archivePrefix={arXiv},
      primaryClass={cs.SE},
      url={https://arxiv.org/abs/2509.16941}, 
}

@misc{yang2026promptsdontsayunderstanding,
      title={What Prompts Don't Say: Understanding and Managing Underspecification in LLM Prompts}, 
      author={Chenyang Yang and Yike Shi and Qianou Ma and Michael Xieyang Liu and Christian Kästner and Tongshuang Wu},
      year={2026},
      eprint={2505.13360},
      archivePrefix={arXiv},
      primaryClass={cs.CL},
      url={https://arxiv.org/abs/2505.13360}, 
}

@misc{larbi2025promptswrongevaluatingcode,
      title={When Prompts Go Wrong: Evaluating Code Model Robustness to Ambiguous, Contradictory, and Incomplete Task Descriptions}, 
      author={Maya Larbi and Amal Akli and Mike Papadakis and Rihab Bouyousfi and Maxime Cordy and Federica Sarro and Yves Le Traon},
      year={2025},
      eprint={2507.20439},
      archivePrefix={arXiv},
      primaryClass={cs.SE},
      url={https://arxiv.org/abs/2507.20439}, 
}

@misc{laban2025llmslostmultiturnconversation,
      title={LLMs Get Lost In Multi-Turn Conversation}, 
      author={Philippe Laban and Hiroaki Hayashi and Yingbo Zhou and Jennifer Neville},
      year={2025},
      eprint={2505.06120},
      archivePrefix={arXiv},
      primaryClass={cs.CL},
      url={https://arxiv.org/abs/2505.06120}, 
}

@misc{swerebenchv2,
      title={SWE-rebench V2: Language-Agnostic SWE Task Collection at Scale}, 
      author={Ibragim Badertdinov and Maksim Nekrashevich and Anton Shevtsov and Alexander Golubev},
      year={2026},
      eprint={2602.23866},
      archivePrefix={arXiv},
      primaryClass={cs.SE},
      url={https://arxiv.org/abs/2602.23866}, 
}

@inproceedings{sliwerski2005szz,
      title={When Do Changes Induce Fixes?},
      author={{\'S}liwerski, Jacek and Zimmermann, Thomas and Zeller, Andreas},
      booktitle={Proceedings of the 2005 International Workshop on Mining Software Repositories (MSR)},
      pages={1--5},
      year={2005},
      doi={10.1145/1083142.1083147},
}

@misc{swebench,
      title={SWE-bench: Can Language Models Resolve Real-World GitHub Issues?}, 
      author={Carlos E. Jimenez and John Yang and Alexander Wettig and Shunyu Yao and Kexin Pei and Ofir Press and Karthik Narasimhan},
      year={2024},
      eprint={2310.06770},
      archivePrefix={arXiv},
      primaryClass={cs.CL},
      url={https://arxiv.org/abs/2310.06770}, 
}

@misc{activeswebench,
      title={Active-SWE: Benchmarking Coding Agents for Proactive Bug Fixing without Issue Reports}, 
      author={Haobin Li and Ping Deng and Weizhong Qian and Liang Jiang and Zhenyu Huang and Mouxing Yang and Xi Peng},
      year={2026},
      eprint={2608.04682},
      archivePrefix={arXiv},
      primaryClass={cs.SE},
      url={https://arxiv.org/abs/2608.04682}, 
}

@misc{securevibebench,
      title={SecureVibeBench: Benchmarking Secure Vibe Coding of AI Agents via Reconstructing Vulnerability-Introducing Scenarios}, 
      author={Junkai Chen and Huihui Huang and Yunbo Lyu and Junwen An and Jieke Shi and Chengran Yang and Ting Zhang and Haoye Tian and Yikun Li and Zhenhao Li and Xin Zhou and Xing Hu and David Lo},
      year={2026},
      eprint={2509.22097},
      archivePrefix={arXiv},
      primaryClass={cs.SE},
      url={https://arxiv.org/abs/2509.22097}, 
}

@misc{repoaudit,
      title={RepoAudit: An Autonomous LLM-Agent for Repository-Level Code Auditing}, 
      author={Jinyao Guo and Chengpeng Wang and Xiangzhe Xu and Zian Su and Xiangyu Zhang},
      year={2025},
      eprint={2501.18160},
      archivePrefix={arXiv},
      primaryClass={cs.SE},
      url={https://arxiv.org/abs/2501.18160}, 
}

@misc{livekbench,
      title={Outrunning LLM Cutoffs: A Live Kernel Crash Resolution Benchmark for All}, 
      author={Chenxi Huang and Alex Mathai and Feiyang Yu and Aleksandr Nogikh and Petros Maniatis and Franjo Ivančić and Eugene Wu and Kostis Kaffes and Junfeng Yang and Baishakhi Ray},
      year={2026},
      eprint={2602.02690},
      archivePrefix={arXiv},
      primaryClass={cs.SE},
      url={https://arxiv.org/abs/2602.02690}, 
}

@misc{swereviewbench,
      title={SWE-Review: Closing the Loop on Issue Resolution with Agentic Code Review}, 
      author={Ruoyu Wang and Jierun Chen and Shaowei Wang and Chaofan Tao and Sidi Yang and Yuxin Jiang and Kim-Hui Yap and Lifeng Shang and Xiaohui Li and Haoli Bai},
      year={2026},
      eprint={2607.06065},
      archivePrefix={arXiv},
      primaryClass={cs.SE},
      url={https://arxiv.org/abs/2607.06065}, 
}

@misc{cvebench,
      title={CVE-Bench: A Benchmark for AI Agents' Ability to Exploit Real-World Web Application Vulnerabilities}, 
      author={Yuxuan Zhu and Antony Kellermann and Dylan Bowman and Philip Li and Akul Gupta and Adarsh Danda and Richard Fang and Conner Jensen and Eric Ihli and Jason Benn and Jet Geronimo and Avi Dhir and Sudhit Rao and Kaicheng Yu and Twm Stone and Daniel Kang},
      year={2025},
      eprint={2503.17332},
      archivePrefix={arXiv},
      primaryClass={cs.CR},
      url={https://arxiv.org/abs/2503.17332}, 
}

@misc{cybergym,
      title={CyberGym: Evaluating AI Agents' Real-World Cybersecurity Capabilities at Scale}, 
      author={Zhun Wang and Tianneng Shi and Jingxuan He and Matthew Cai and Jialin Zhang and Dawn Song},
      year={2026},
      eprint={2506.02548},
      archivePrefix={arXiv},
      primaryClass={cs.CR},
      url={https://arxiv.org/abs/2506.02548}, 
}

@ARTICLE{kamei,
  author={Kamei, Yasutaka and Shihab, Emad and Adams, Bram and Hassan, Ahmed E. and Mockus, Audris and Sinha, Anand and Ubayashi, Naoyasu},
  journal={IEEE Transactions on Software Engineering}, 
  title={A large-scale empirical study of just-in-time quality assurance}, 
  year={2013},
  volume={39},
  number={6},
  pages={757-773},
  doi={10.1109/TSE.2012.70}}

@article{zhao,
author = {Zhao, Yunhua and Damevski, Kostadin and Chen, Hui},
title = {A Systematic Survey of Just-in-Time Software Defect Prediction},
year = {2023},
issue_date = {October 2023},
publisher = {Association for Computing Machinery},
address = {New York, NY, USA},
volume = {55},
number = {10},
issn = {0360-0300},
url = {https://doi.org/10.1145/3567550},
doi = {10.1145/3567550},
journal = {ACM Comput. Surv.},
month = feb,
articleno = {201},
numpages = {35}
}

@misc{ticketlevelbug,
      title={Anticipating Bugs: Ticket-Level Bug Prediction and Temporal Proximity Effects}, 
      author={Daniele La Prova and Emanuele Gentili and Davide Falessi},
      year={2025},
      eprint={2506.14290},
      archivePrefix={arXiv},
      primaryClass={cs.SE},
      url={https://arxiv.org/abs/2506.14290}, 
}

@misc{redef,
      title={ReDef: Do Code Language Models Truly Understand Code Changes for Just-in-Time Software Defect Prediction?}, 
      author={Doha Nam and Taehyoun Kim and Duksan Ryu and Jongmoon Baik},
      year={2026},
      eprint={2509.09192},
      archivePrefix={arXiv},
      primaryClass={cs.SE},
      url={https://arxiv.org/abs/2509.09192}, 
}

@misc{codeflowlm,
      title={CodeFlowLM: Incremental Just-In-Time Defect Prediction with Pretrained Language Models and Exploratory Insights into Defect Localization}, 
      author={Monique Louise Monteiro and George G. Cabral and Adriano L. I. OLiveira},
      year={2025},
      eprint={2512.00231},
      archivePrefix={arXiv},
      primaryClass={cs.SE},
      url={https://arxiv.org/abs/2512.00231}, 
}

@misc{glm53,
    title={GLM-5.3-Flash: Frontier Intelligence, Flash Cost},
    author={Z.ai},
    url={https://z.ai/blog/glm-5.3-flash}
}

@article{zhou2025tom,
  title={Tom-swe: User mental modeling for software engineering agents},
  author={Zhou, Xuhui and Chen, Valerie and Wang, Zora Zhiruo and Neubig, Graham and Sap, Maarten and Wang, Xingyao},
  journal={arXiv preprint arXiv:2510.21903},
  year={2025}
}

@misc{codex,
    title={Introducing Codex},
    author={OpenAI Team},
    url={https://openai.com/index/introducing-codex/}
}

@misc{claudecode,
    title={Claude Code},
    author={Anthropic Team},
    url={https://claude.com/product/claude-code}
}

@inproceedings{yang2024sweagent,
  title={{SWE}-agent: Agent-Computer Interfaces Enable Automated Software Engineering},
  author={John Yang and Carlos E Jimenez and Alexander Wettig and Kilian Lieret and Shunyu Yao and Karthik R Narasimhan and Ofir Press},
  booktitle={The Thirty-eighth Annual Conference on Neural Information Processing Systems},
  year={2024},
  url={https://arxiv.org/abs/2405.15793}
}

@misc{deepseekv4,
      title={DeepSeek-V4: Towards Highly Efficient Million-Token Context Intelligence}, 
      author={DeepSeek-AI and Anyi Xu and Bangcai Lin and Bing Xue and Bingxuan Wang and Bingzheng Xu and Bochao Wu and Bowei Zhang and Chaofan Lin and Chen Dong and Chenchen Ling and Chengda Lu and Chenggang Zhao and Chengqi Deng and Chengyu Hou and Chenhao Xu and Chenze Shao and Chong Ruan and Conner Sun and Damai Dai and Daya Guo and Dejian Yang and Deli Chen and Donghao Li and Dongjie Ji and Erhang Li and Fang Wei and Fangyun Lin and Fangzhou Yuan and Feiyu Xia and Fucong Dai and Guangbo Hao and Guanting Chen and Guoai Cao and Guolai Meng and Guowei Li and Han Yu and Han Zhang and Hanwei Xu and Hao Li and Haofen Liang and Haoling Zhang and Haoming Luo and Haoran Wei and Haotian Yuan and Haowei Zhang and Haowen Luo and Haoyu Chen and Haozhe Ji and Hengqing Zhang and Honghui Ding and Hongxuan Tang and Huanqi Cao and Huazuo Gao and Hui Qu and Hui Zeng and J Yang and JQ Zhu and Jia Luo and Jia Song and Jia Yu and Jialiang Huang and Jialu Cai and Jian Liang and Jiangting Zhou and Jiasheng Ye and Jiashi Li and Jiaxin Xu and Jiewen Hu and Jieyu Yang and Jin Chen and Jin Yan and Jingchang Chen and Jingli Zhou and Jingting Xiang and Jingyang Yuan and Jingyuan Cheng and Jingzi Zhou and Jinhua Zhu and Jiping Yu and Joseph Sun and Jun Ran and Junguang Jiang and Junjie Qiu and Junlong Li and Junmin Zheng and Junxiao Song and Kai Dong and Kaige Gao and Kang Guan and Kexing Zhou and Kezhao Huang and Kuai Yu and Lean Wang and Lecong Zhang and Lei Wang and Leyi Xia and Li Zhang and Liang Zhao and Lihua Guo and Lingxiao Luo and Linwang Ma and Linyan Zhu and Litong Wang and Liyu Cai and Liyue Zhang and Longhao Chen and MS Di and MY Xu and Max Mei and Miaojun Wang and Mingchuan Zhang and Minghua Zhang and Minghui Tang and Mingming Li and Mingxu Zhou and Minmin Han and Ning Wang and Panpan Huang and Panpan Wang and Peixin Cong and Peiyi Wang and Peng Zhang and Qiancheng Wang and Qihao Zhu and Qingyang Li and Qinyu Chen and Qiushi Du and Qiwei Jiang and Rui Tian and Ruifan Xu and Ruijie Lu and Ruiling Xu and Ruiqi Ge and Ruisong Zhang and Ruizhe Pan and Runji Wang and Runqian Chen and Runqiu Yin and Runxin Xu and Ruomeng Shen and Ruoyu Zhang and Ruyi Chen and SH Liu and Shanghao Lu and Shangmian Sun and Shangyan Zhou and Shanhuang Chen and Shaofei Cai and Shaoheng Nie and Shaoqing Wu and Shaoyuan Chen and Shengding Hu and Shengyu Liu and Shiqiang Hu and Shirong Ma and Shiyu Wang and Shuiping Yu and Shunfeng Zhou and Shuting Pan and Shuying Yu and Songyang Zhou and Tao Ni and Tao Yun and Tian Jin and Tian Pei and Tian Ye and Tianle Lin and Tianran Ji and Tianyi Cui and Tianyuan Yue and Tingting Yu and Tun Wang and W Zhang and WL Xiao and Wangding Zeng and Wei An and Weilin Zhao and Wen Liu and Wenfeng Liang and Wenjie Pang and Wenjing Luo and Wenjing Yao and Wenjun Gao and Wenkai Yang and Wenlve Huang and Wenqing Hou and Wentao Zhang and Wenting Ma and Xi Gao and Xiang He and Xiangwen Wang and Xianzu Wang and Xiao Bi and Xiaodong Liu and Xiaohan Wang and Xiaokang Chen and Xiaokang Zhang and Xiaotao Nie and Xiaowen Sun and Xiaoxiang Wang and Xin Cheng and Xin Liu and Xin Xie and Xingchao Liu and Xingchen Liu and Xingkai Yu and Xingyou Li and Xinyu Yang and Xinyu Zhang and Xu Chen and Xuanyu Wang and Xuecheng Su and Xueyin Chen and Xuheng Lin and Xuwei Fu and YC Yan and YQ Wang and YW Ma and Yanfeng Luo and Yang Zhang and Yanhong Xu and Yanru Ma and Yanwen Huang and Yao Li and Yao Li and Yao Xu and Yao Zhao and Yaofeng Sun and Yaohui Wang and Yi Qian and Yi Shao and Yi Yu and Yichao Zhang and Yifan Ding and Yifan Shi and Yijia Wu and Yiliang Xiong and Yiling Ma and Ying He and Ying Tang and Ying Zhou and Yingjia Luo and Yinmin Zhong and Yishi Piao and Yisong Wang and Yixiang Zhang and Yixiao Chen and Yixuan Tan and Yixuan Wei and Yiyang Ma and Yiyuan Liu and Yonglun Yang and Yongqiang Guo and Yongtong Wu and Yu Wu and YuKun Li and Yuan Cheng and Yuan Ou and Yuanfan Xu and Yuanhao Li and Yuduan Wang and Yuehan Yang and Yuer Xu and Yuhan Wu and Yuhao Meng and Yuheng Zou and Yukun Zha and Yunfan Xiong and Yupeng Chen and Yuping Lin and Yuqian Cao and Yuqian Wang and Yushun Zhang and Yuting Yan and Yutong Lin and Yuxian Gu and Yuxiang Luo and Yuxiang You and Yuxuan Liu and Yuxuan Zhou and Yuyang Zhou and Yuzhen Huang and ZF Wu and Zehao Wang and Zehua Zhao and Zehui Ren and Zekai Zhang and Zhangli Sha and Zhe Fu and Zhe Ju and Zhean Xu and Zhenda Xie and Zhengyan Zhang and Zheren Gao and Zhewen Hao and Zhibin Gou and Zhicheng Ma and Zhigang Yan and Zhihong Shao and Zhixian Huang and Zhixuan Chen and Zhiyu Wu and Zhizhou Ren and Zhongyu Wu and Zhuoshu Li and Zhuping Zhang and Zian Xu and Zihao Wang and Zihua Qu and Zihui Gu and Zijia Zhu and Zilin Li and Zipeng Zhang and Ziwei Xie and Ziyi Gao and Ziyi Wan and Zizheng Pan and Zongqing Yao},
      year={2026},
      eprint={2606.19348},
      archivePrefix={arXiv},
      primaryClass={cs.CL},
      url={https://arxiv.org/abs/2606.19348}, 
}

@inproceedings {detectSilentFailures,
author = {Chang Lou and Dimas Shidqi Parikesit and Yujin Huang and Zhewen Yang and Senapati Diwangkara and Yuzhuo Jing and Achmad Imam Kistijantoro and Ding Yuan and Suman Nath and Peng Huang},
title = {Deriving Semantic Checkers from Tests to Detect Silent Failures in Production Distributed Systems},
booktitle = {19th USENIX Symposium on Operating Systems Design and Implementation (OSDI 25)},
year = {2025},
isbn = {978-1-939133-47-2},
address = {Boston, MA},
pages = {19--38},
url = {https://www.usenix.org/conference/osdi25/presentation/lou},
publisher = {USENIX Association},
month = jul
}

@inproceedings {demystifyingSilentErrors,
author = {Chang Lou and Yuzhuo Jing and Peng Huang},
title = {Demystifying and Checking Silent Semantic Violations in Large Distributed Systems},
booktitle = {16th USENIX Symposium on Operating Systems Design and Implementation (OSDI 22)},
year = {2022},
isbn = {978-1-939133-28-1},
address = {Carlsbad, CA},
pages = {91--107},
url = {https://www.usenix.org/conference/osdi22/presentation/lou-demystifying},
publisher = {USENIX Association},
month = jul
}

@ARTICLE{odc,
  author={Chillarege, R. and Bhandari, I.S. and Chaar, J.K. and Halliday, M.J. and Moebus, D.S. and Ray, B.K. and Wong, M.-Y.},
  journal={IEEE Transactions on Software Engineering}, 
  title={Orthogonal defect classification-a concept for in-process measurements}, 
  year={1992},
  volume={18},
  number={11},
  pages={943-956},
  doi={10.1109/32.177364}}

@misc{lyu2026agentszzteachingllmagent,
      title={AgentSZZ: Teaching the LLM Agent to Play Detective with Bug-Inducing Commits}, 
      author={Yunbo Lyu and Jieke Shi and Hong Jin Kang and Ratnadira Widyasari and Junda He and Yuqing Niu and Chengran Yang and Junkai Chen and Zhou Yang and Julia Lawall and David Lo},
      year={2026},
      eprint={2604.02665},
      archivePrefix={arXiv},
      primaryClass={cs.SE},
      url={https://arxiv.org/abs/2604.02665}, 
}

@article{10.1145/3728885,
author = {Tang, Lingxiao and Liu, Jiakun and Liu, Zhongxin and Yang, Xiaohu and Bao, Lingfeng},
title = {LLM4SZZ: Enhancing SZZ Algorithm with Context-Enhanced Assessment on Large Language Models},
year = {2025},
issue_date = {July 2025},
publisher = {Association for Computing Machinery},
address = {New York, NY, USA},
volume = {2},
number = {ISSTA},
url = {https://doi.org/10.1145/3728885},
doi = {10.1145/3728885},
journal = {Proc. ACM Softw. Eng.},
month = jun,
articleno = {ISSTA016},
numpages = {23}
}

@misc{cao2026masszzmultiagenticszzalgorithm,
      title={MAS-SZZ: Multi-Agentic SZZ Algorithm for Vulnerability-Inducing Commit Identification}, 
      author={Sicong Cao and Jinxuan Xu and Le Yu and Jing Yang and Xingwei Lin and Linlin Zhu and Fu Xiao},
      year={2026},
      eprint={2604.24398},
      archivePrefix={arXiv},
      primaryClass={cs.CR},
      url={https://arxiv.org/abs/2604.24398}, 
}

@misc{swechain,
      title={SWE-Chain: Benchmarking Coding Agents on Chained Release-Level Package Upgrades}, 
      author={Man Ho Lam and Chaozheng Wang and Hange Liu and Jingyu Xiao and Haau-sing Li and Jen-tse Huang and Terry Yue Zhuo and Michael R. Lyu},
      year={2026},
      eprint={2605.14415},
      archivePrefix={arXiv},
      primaryClass={cs.SE},
      url={https://arxiv.org/abs/2605.14415}, 
}

@misc{programbench,
      title={ProgramBench: Can Language Models Rebuild Programs From Scratch?}, 
      author={John Yang and Kilian Lieret and Jeffrey Ma and Parth Thakkar and Dmitrii Pedchenko and Sten Sootla and Emily McMilin and Pengcheng Yin and Rui Hou and Gabriel Synnaeve and Diyi Yang and Ofir Press},
      year={2026},
      eprint={2605.03546},
      archivePrefix={arXiv},
      primaryClass={cs.SE},
      url={https://arxiv.org/abs/2605.03546}, 
}

@misc{forecastbench,
      title={ForecastBench: A Dynamic Benchmark of AI Forecasting Capabilities}, 
      author={Ezra Karger and Houtan Bastani and Chen Yueh-Han and Zachary Jacobs and Danny Halawi and Fred Zhang and Philip E. Tetlock},
      year={2025},
      eprint={2409.19839},
      archivePrefix={arXiv},
      primaryClass={cs.LG},
      url={https://arxiv.org/abs/2409.19839}, 
}

@misc{mirai,
      title={MIRAI: Evaluating LLM Agents for Event Forecasting}, 
      author={Chenchen Ye and Ziniu Hu and Yihe Deng and Zijie Huang and Mingyu Derek Ma and Yanqiao Zhu and Wei Wang},
      year={2024},
      eprint={2407.01231},
      archivePrefix={arXiv},
      primaryClass={cs.CL},
      url={https://arxiv.org/abs/2407.01231}, 
}

@inproceedings{arethesebugsreallynormal,
author = {Saha, Ripon K. and Lawall, Julia and Khurshid, Sarfraz and Perry, Dewayne E.},
title = {Are these bugs really "normal"?},
year = {2015},
isbn = {9780769555942},
publisher = {IEEE Press},
booktitle = {Proceedings of the 12th Working Conference on Mining Software Repositories},
pages = {258–268},
numpages = {11},
location = {Florence, Italy},
series = {MSR '15}
}

@misc{weave,
      title={WEAVE: Weaving Real-World Code Evolution into Dependency-Structured Tasks for Benchmarking and Training Code Agents}, 
      author={Weihao Li and Zihang Xu and Zhongqiu Wang and Songwei Yu and Wenqiang Zhu and Yuechen Luo and Jingxuan Xu and Yanan Wu and Haoyang Huang and Aerial Deng},
      year={2026},
      eprint={2610.22039},
      archivePrefix={arXiv},
      primaryClass={cs.SE},
      url={https://arxiv.org/abs/2610.22039}, 
}

@misc{davinciagency,
      title={daVinci-Agency: Unlocking Long-Horizon Agency Data-Efficiently}, 
      author={Mohan Jiang and Dayuan Fu and Junhao Shi and Ji Zeng and Weiye Si and Keyu Li and Xuefeng Li and Yang Xiao and Wenjie Li and Dequan Wang and Pengfei Liu},
      year={2026},
      eprint={2602.02619},
      archivePrefix={arXiv},
      primaryClass={cs.LG},
      url={https://arxiv.org/abs/2602.02619}, 
}

@InProceedings{swemilestone,
  title = 	 {{SWE}-Milestone: Evaluating {AI} Agents on Continuous Software Evolution},
  author =       {Deng, Gangda and Chen, Zhaoling and Yu, Zhongming and Fan, Haoyang and Liu, Yuhong and Yang, Yuxin and Parikh, Dhruv and Kannan, Rajgopal and Cong, Le and Wang, Mengdi and Zhang, Qian and Prasanna, Viktor and Tang, Xiangru and Wang, Xingyao},
  booktitle = 	 {Proceedings of the 43rd International Conference on Machine Learning},
  pages = 	 {23853--23890},
  year = 	 {2026},
  editor = 	 {Zhang, Tong and Dudik, Miroslav and Jaggi, Martin and Agarwal, Alekh and Li, Sharon and Schuurmans, Dale and Zhu, Jerry and Berkenkamp, Felix and Dong, Hanze and Bietti, Alberto},
  volume = 	 {306},
  series = 	 {Proceedings of Machine Learning Research},
  month = 	 {06--11 Jul},
  publisher =    {PMLR},
  url = 	 {https://proceedings.mlr.press/v306/deng26d.html}
}

@misc{chainswe,
      title={ChainSWE: Benchmarking Coding Agents on Multi-Bug Software Maintenance}, 
      author={Qirui Jin and Lingching Tung and Kenan Li and Qiyang Shi and Yushi She and Huanzhong Jia and Harrison Zhao and Kejing Xia and Zhenbang Du and Jiaxin Pei and Zhenyu Zhang and Zhen Qi and Yuyan Duan and Wenke Lee and Zijian Jin},
      year={2026},
      eprint={2607.02606},
      archivePrefix={arXiv},
      primaryClass={cs.SE},
      url={https://arxiv.org/abs/2607.02606}, 
}

@misc{sweci,
      title={SWE-CI: Evaluating Agent Capabilities in Maintaining Codebases via Continuous Integration}, 
      author={Jialong Chen and Xander Xu and Hu Wei and Chuan Chen and Bing Zhao},
      year={2026},
      eprint={2603.03823},
      archivePrefix={arXiv},
      primaryClass={cs.SE},
      url={https://arxiv.org/abs/2603.03823}, 
}

@article{slopcodebench,
  title={SlopCodeBench: Benchmarking How Coding Agents Degrade Over Long-Horizon Iterative Tasks},
  author={Orlanski, Gabriel and Roy, Devjeet and Yun, Alexander and Shin, Changho and Gu, Alex and Ge, Albert and Adila, Dyah and Roberts, Nicholas and Sala, Frederic and Albarghouthi, Aws},
  journal={arXiv preprint arXiv:2603.24755},
  year={2026}
}

@misc{codethread,
      title={Is Agent Code Less Maintainable Than Human Code?}, 
      author={Shaswat Patel and Betty Li Hou and Arun Purohit and Kai Xu and Jane Pan and He He and Valerie Chen},
      year={2026},
      eprint={2606.21804},
      archivePrefix={arXiv},
      primaryClass={cs.SE},
      url={https://arxiv.org/abs/2606.21804}, 
}

@misc{swesweep,
  title  = {{SWE-sweep}: Can Agents Autonomously Find and Fix Bugs?},
  author = {Kilian Lieret and Jeffrey Jian Ma and Rahul Kindi and
            Yuxiang Wei and Jeremy Ma and Sten Sootla and
            Parth Thakkar and Chao Beyond Zhou and Pengcheng Yin and
            Rui Hou and Ofir Press and John Yang},
  year   = {2026},
  note   = {Preprint}
}
\bibliographystyle{iclr2027_conference}

\clearpage          
\appendix
\setcounter{table}{0}
\setcounter{figure}{0}
\renewcommand{\thetable}{\thesection.\arabic{table}}
\renewcommand{\thefigure}{\thesection.\arabic{figure}}
\counterwithin{table}{section}
\counterwithin{figure}{section}
\section*{Appendix}

\appendix
\section{Benchmark Construction}
\label{appendix:benchmark_construction}

\subsection{Identify valid foresight pairs}\label{appendix:step_2}

\subsubsection{PR Pair validity judge}\label{appendix:pair_validity}

\begin{promptbox}{Pair Validity Template}
  
    \pbhead{Evidence}
  
  Repository: \textbf{[Repo]}
  
  PR1's task (what the earlier change set out to do): \textless{}\textless{}\textless{} \textbf{[Pr1 Task]} \textgreater{}\textgreater{}\textgreater{}
  
  PR2's description (the fix for the missed edge case): \textless{}\textless{}\textless{} \textbf{[Pr2 Description]} \textgreater{}\textgreater{}\textgreater{}
  
  PR2's fix patch (the gold anticipation fix): \textless{}\textless{}\textless{} \textbf{[Fix Patch]} \textgreater{}\textgreater{}\textgreater{}
  
  PR2's test patch (the HELD-OUT tests --- fail before PR2, pass after): \textless{}\textless{}\textless{} \textbf{[Test Patch]} \textgreater{}\textgreater{}\textgreater{}
  
  Human discussion (PR + linked issues/tickets): \textless{}\textless{}\textless{} \textbf{[Discussion]} \textgreater{}\textgreater{}\textgreater{}
  
  Objective evidence (gap, held-out test ids, changelog lines): \textless{}\textless{}\textless{} \textbf{[Evidence]} \textgreater{}\textgreater{}\textgreater{}
  
  \pbhead{Weigh everything together}
  
  No single source is authoritative --- piece the whole story together from ALL of the above. Different facts surface in different places: a false-positive vs a genuine miss is often clearest from the fix patch + tests; a revert/softening or
  the reason for it often only appears in the discussion or changelog; the descriptions frame intent. Synthesize, then extract the two facts.
  
  \pbhead{Fact 1 --- eligibility: is PR2 a genuine, anticipatable missed-edge-case?}
  
  \begin{pbitemize}
    \item "clean" --- PR2 fixes a genuine BUG: an edge case of PR1's behavior that produced a defect (wrong output, crash, silent wrong result) on some input. A real user-facing defect an agent could have anticipated. This is the only KEEP
  class. SECURITY fixes count as clean: a change that closes an attack (weak/predictable randomness, injection, XSS, path traversal, auth/authorization bypass, CRLF/header injection, SSRF, ReDoS/DoS) is a real bug even if there is NO
  observable defect for legitimate users --- security bugs are silent to the victim.
    \item "not\_a\_bug" --- no user-facing defect AND not a security fix: pure refactor, rename/alias, docstring/docs, deprecation bookkeeping, internal hygiene, type-only change. (Say what non-bug purpose it served.) NB: "no observable
  defect for legitimate users" does NOT imply not\_a\_bug --- if the change closes an attack, it is "clean" (see above).
    \item "new\_feature" --- PR2 ADDS new capability rather than fixing a defect in existing behavior; the held-out test exercises a NEW API/feature, not a missed edge case of what PR1 did.
    \item "unpredictable" --- PR2 REVERTS or SOFTENS a change PR1 made INTENTIONALLY, for SOCIAL/POLICY reasons (community pushback, "too disruptive", deprecation timing) --- not an objective technical defect. An agent cannot foresee a
  maintainer's later change of heart; the reversal reason usually appears in the discussion or changelog. (BUT: if PR2 reverts because PR1's intentional change had an OBJECTIVE defect visible from spec/code --- e.g. its stricter validation
  rejected valid inputs --- that is "clean", not unpredictable.)
    \item "weak" --- a real but trivial/borderline defect, or you are genuinely unsure it is a meaningful edge case.
  \end{pbitemize}
  
  \pbhead{Fact 2 --- task validity: are PR2's held-out tests a well-formed, FAIR task?}
  
  "A" if none of the issues below apply. Otherwise the FIRST applicable Bxx code. These flag when passing the held-out tests requires more (or other) than solving the stated edge case --- read the test patch closely.
  
  {\footnotesize
  \begin{tabularx}{\linewidth}{@{}llX@{}}
  \toprule
    \textbf{Code} & \textbf{Issue} & \textbf{Signal} \\
  \midrule
    B1 & TEST\_SUITE\_COUPLING & tests span unrelated modules; a correct fix could fail on unrelated regressions \\
    B2 & IMPLICIT\_NAMING & tests assert specific names / signatures / error-message strings / output values that come from the patch, not derivable from the problem \\
    B3 & EXTERNAL\_DEPENDENCY & essential info lives in external URLs not quoted in the issue/PR \\
    B4 & AMBIGUOUS\_SPEC & missing expected behavior, repro, or acceptance criteria \\
    B5 & PATCH\_ARTIFACTS & the fix patch bundles UNRELATED changes that the held-out tests nonetheless require \\
    B6 & IMPLICIT\_KNOWLEDGE & tests assert a specific approach/internals not stated or inferrable from the repo \\
    B7 & INLINE\_TEST & a test is inlined in the fix (source) patch rather than the test patch \\
    B8 & VISUAL\_REQUIREMENT & task needs images/screenshots/visual context (incl.\ markdown images) not readable as text \\
    B9 & NON\_ENGLISH\_TEXT & the problem statement is partially/fully non-English \\
    B10 & TEXT\_FILE\_TESTS & the test patch only changes plain-text fixtures (\texttt{.txt}/\texttt{.csv}/\dots), not code tests \\
    B11 & MULTI\_ISSUE\_PR & PR2 fixes multiple independent issues; ambiguous which change the tests require \\
  \bottomrule
  \end{tabularx}}
  
  \pbhead{Output (JSON only, NO keep/drop decision)}
    
  \begin{lstlisting}
  { 
    "eligibility": {"value": "clean|not_a_bug|new_feature|unpredictable|weak", "evidence": "short quote"},
    "validity":    {"code": "A|B1|B2|B3|B4|B5|B6|B7|B8|B9|B10|B11", "evidence": "short quote or 'none'"},
    "reasoning":   "2-3 sentences: what PR2 fixes, why that eligibility, and any validity concern",
    "confidence":  0.0
  }
  \end{lstlisting}
\end{promptbox}

\subsubsection{SZZ-attribution judge}\label{appendix:szz_attribution}
  
\begin{promptbox}{SZZ Attribution Template}
  You are auditing an automated bug-attribution claim in the repository \textbf{[Repo]}. 
  
  An automated tool claims that an earlier pull request (PR1) introduced the defect that a later pull request (PR2) fixes. It reached that claim \textbf{mechanically}: it took the lines PR2 deleted or modified, ran \texttt{git blame} on
  them, and followed the blame back to PR1. 
  
  That procedure is frequently wrong. Blame finds whoever last \emph{touched} a line, which is not the same as whoever \emph{broke} it. Your job is to decide whether the causal claim actually holds.
  
  \textbf{Treat the claim as a hypothesis, not a fact.} Nothing below is evidence that PR1 is the true introducer --- PR1 is simply where the blame landed. Do not assume the link is correct because this prompt describes PR1 as the
  introducer. Judge it from the code.
  
  \pbhead{Evidence}
  
  Repository: \textbf{[Repo]}
  
  PR1 --- what the earlier change set out to do: \textless{}\textless{}\textless{} \textbf{[Pr1 Task]} \textgreater{}\textgreater{}\textgreater{}
  
  PR1's diff. \textbf{[Pr1 Scope Note]} \textless{}\textless{}\textless{} \textbf{[Pr1 Diff]} \textgreater{}\textgreater{}\textgreater{} 
  
  \textbf{The connecting lines} --- lines PR2 deleted or modified, which \texttt{git blame} traces to a commit inside PR1. These are the entire basis of the tool's claim: \textless{}\textless{}\textless{} \textbf{[Blamed Lines]}
  \textgreater{}\textgreater{}\textgreater{}
  
  \textbf{Changes to those files since PR1} --- \textbf{[Intervening N]} commit(s) touched the connecting lines' file(s) between PR1 and PR2. If one of these introduced the condition PR2 fixes, PR1 is not the origin:
  \textless{}\textless{}\textless{} \textbf{[Intervening]} \textgreater{}\textgreater{}\textgreater{}
  
  PR2 --- the fix: \textless{}\textless{}\textless{} \textbf{[Pr2 Description]} \textgreater{}\textgreater{}\textgreater{}
  
  PR2's fix patch: \textless{}\textless{}\textless{} \textbf{[Fix Patch]} \textgreater{}\textgreater{}\textgreater{}
  
  PR2's test patch (the tests that fail before the fix and pass after): \textless{}\textless{}\textless{} \textbf{[Test Patch]} \textgreater{}\textgreater{}\textgreater{}
  
  Human discussion (PR comments and linked issues, both PRs): \textless{}\textless{}\textless{} \textbf{[Discussion]} \textgreater{}\textgreater{}\textgreater{}
  
  Issue and PR numbers have been replaced with \texttt{\textless{}PR-Number\textgreater{}} throughout. That is deliberate --- decide from the code and the described behaviour, not from cross-references.
  
  \pbhead{How to decide}
    
  Everything turns on one question: \textbf{what did PR1's diff actually DO to the connecting lines?} Work through it in this order.

    \begin{pbenumerate}
    \item \textbf{Find the defect.} Read PR2's fix patch and its tests. What input or state produced the wrong behaviour, and what does the fix change?
    \item \textbf{Find that same logic in PR1's diff.} Look at the connecting lines and locate them in PR1's change.
    \item \textbf{Ask what PR1 did to them.} Did PR1 \emph{write} the flawed logic --- the wrong condition, the missing guard, the bad default --- as new lines of its own? Did PR1 write those lines \emph{correctly} for what existed at the
  time, with PR2 now handling a case PR1's change never covered or created? Did PR1 merely \emph{move} the lines --- a rename, a file move, reformatting, an import or type-annotation change, a lint or codemod pass? A PR1 that swept a large
  number of files is a strong hint of this.
    \item \textbf{Check the change list against the defect before you accept.} You named the defect in step 1. Now read the commit subjects in the change list above and ask a single question of each: \emph{does this describe work on the same
  behaviour, subsystem or symptom as the defect?} If any subject does, you may not choose A without saying why that commit is not the origin --- name it in \texttt{later\_change} and give the reason. This is the most common way this audit
  goes wrong: the culprit is sitting in the change list, described plainly, and the judge accepts PR1 without ever comparing the two. Scan subjects only; you are not being asked to reason about their contents.
  \end{pbenumerate}
  
  A defect can have more than one contributing change. Judge whether PR1 is a \textbf{credible cause of this specific defect}, not whether it is the only possible one.

  \pbhead{Fact --- attribution}
  
  Choose exactly one code. The illustrations below are schematic, not real cases.
  
  \begin{pbitemize}
    \item "A" --- AUTHORED. PR1's own diff introduces the logic PR2 corrects. The flawed condition, missing case, wrong default or absent guard is visible among the lines PR1 added or changed. \textbf{This is the only code meaning the
  attribution HOLDS.} \emph{Example:} PR1 adds \texttt{if timeout:} to apply a socket timeout. PR2 changes it to \texttt{if timeout is not None}, because \texttt{timeout=0} was silently ignored. PR1 wrote the faulty condition -\textgreater{}
  A.
    \item "C1" --- STALE\_ORIGIN. PR1 authored the connecting lines, but they were correct as written and the defect arrived later. To use C1 you must \textbf{name the specific later change} from the list above (or show the flawed logic
  predates PR1 entirely). This is the most common way blame misleads: it reaches back to whoever originally wrote a line that only later became inadequate. \emph{Example:} PR1 adds \texttt{def decode(self, data):} as a synchronous method,
  correct for every caller then. The change list shows a later commit ``make decode() async''; PR2 restores the sync API. A specific later change is nameable -\textgreater{} C1.
  \end{pbitemize}
  
  \textbf{A long change list is not itself C1.} Many commits may have touched the file without touching the behaviour PR2 fixes. Conversely, an empty change list is strong evidence \emph{against} C1: nothing happened in between, so PR1's
  code is what PR2 is correcting.
  
  \begin{pbitemize}
    \item "C2" --- CARRIED. PR1 only relocated or reshaped the lines --- a move between files, a rename, reformatting, import reorder, type annotation, or a lint/codemod sweep. The behaviour predates PR1; PR1 carried it across unchanged, so
  blame landed on PR1 by accident. A PR1 that swept a large number of files is a strong hint. \emph{Example:} PR1 moves \texttt{\_normalize()} from \texttt{utils.py} to \texttt{text.py} with its body unchanged, and reformats it. PR2 fixes a
  bug inside \texttt{\_normalize} that was present before the move -\textgreater{} C2.
    \item "C3" --- UNCLEAR. The evidence does not let you tell: PR1's diff does not show enough of the relevant logic, or the defect cannot be located in what you were given.
  \end{pbitemize}
  
  The line between A and C1 is \textbf{not} whether PR1 touched the code --- it did, in both. It is whether PR1's change \emph{created the flaw}, or merely wrote something that was fine at the time. Ask: given only what existed when PR1
  landed, would a careful reviewer have called PR1's code wrong? If yes, A. If PR1's code was reasonable then and something later made it inadequate, C1.
  
  Prefer a definite code when the evidence supports one. Use C3 when it genuinely does not --- not as a hedge.
  
  \pbhead{Output (JSON only)}
  
  Answer the three questions FIRST, then commit to a code. Emit ONE flat JSON object with exactly these six keys --- no nested objects, no extra keys, nothing outside the object.
  
  \begin{lstlisting}
  {
    "defect": "one sentence: what PR2 actually fixes",
    "pr1_role": "one sentence: what PR1 did to the connecting lines",
    "reasoning": "2-3 sentences working through steps 1-3 above",
    "later_change": "for C1: quote the commit from the change list that introduced the defect. Otherwise \"none\".",
    "code": "A|C1|C2|C3",
    "evidence": "short quote from PR1's diff or the connecting lines that decides it"
  } 
  \end{lstlisting}
  
  All six keys are required. Write \texttt{reasoning} before choosing \texttt{code} --- the analysis is what the code follows from, not a justification added afterwards.
\end{promptbox}

\subsection{Construct and validate executable instances}\label{appendix:env_synth_template}
Each instance must be reproduced at four different commits, and a single installation recipe is not sufficient across a repository’s history. Over the time span we mine, projects often move test dependencies between files such as \texttt{setup.cfg}, and \texttt{pyproject.toml}. We therefore generate the environment separately for each commit using an LLM agent, which produces the installation command artifact used by SWE-ReBench V2 \citep{swerebenchv2}. We use DeepSeek-V4-Flash~\citep{deepseekv4} as LLM in a simple harness.

\begin{promptbox}{Installer Agent Template}
\pbsub{System}

\{preamble\}

GOAL: find the commands that install this project and its TEST dependencies, so that this exact command produces a JUnit XML file containing at least one \textless{}testcase\textgreater{}:

\{accept\_cmd\}

\textbf{* FAILING TESTS ARE EXPECTED AND ARE NOT YOUR PROBLEM. }*

You are configuring an ENVIRONMENT, not fixing code. Some of the test files in this working tree were deliberately copied in from a LATER commit, so they test behaviour this commit does not have yet. They are SUPPOSED to fail. A run that reports "4 failed, 128 passed" is a COMPLETE SUCCESS -- the tests executed, which is all we need. Report your config immediately when you see output like that.

You are done as soon as the command above RUNS the tests and writes the XML. You are NOT done only when tests pass, and you must never try to make them pass. Do not read or edit source files under src/, do not investigate an AssertionError, do not downgrade or pin a package to make a test green.

The ONLY failures that mean your job is unfinished are ones that stop tests from RUNNING at all:

\begin{pbitemize}
  \item ModuleNotFoundError / ImportError for a missing dependency
  \item "no tests ran", an empty XML, or a usage error before collection
  \item a collection error caused by an uninstalled pytest plugin
\end{pbitemize}

\pbhead{Response format -- read this twice}

Your response must contain a THOUGHT section and then EXACTLY ONE bash code block holding ONE command (or commands joined with \&\& or ||). Format it exactly as shown:

\textless{}format\_example\textgreater{} THOUGHT: Your reasoning and analysis here

\begin{lstlisting}
your_command_here
\end{lstlisting}

\textless{}/format\_example\textgreater{}

The block MUST be a markdown fence opened with three backticks followed by \texttt{bash} and closed with three backticks. Do NOT use XML-style tags such as \textless{}bash\textgreater{}...\textless{}/bash\textgreater{}. Do NOT emit more than one block. Anything else is unparseable, your command will not run, and the turn is wasted.

Failure to follow these rules will cause your response to be rejected.

I will run the command and reply with its output. Rules: \{toolchain\_rules\}

\begin{pbitemize}
  \item SYSTEM PACKAGES ARE YOURS TO INSTALL. You are root with network access, so if a
\end{pbitemize}

build needs a compiler, a header, a library, or a toolchain the image does not have -- \texttt{apt-get update \&\& apt-get install -y \textless{}pkg\textgreater{}} and carry on. A vendored C or JS dependency may need building before the package will compile (for example a submodule under vendor/ with its own Makefile or package.json); \texttt{git submodule update --init --recursive} and that build are legitimate steps too.

\begin{pbitemize}
  \item EVERY command needed to reproduce the environment must appear in your final
\end{pbitemize}

\texttt{install} list, including apt-get and any vendored build. The list is REPLAYED in a fresh container that has none of your changes -- a step you leave out is a step that will be missing when it matters.

\begin{pbitemize}
  \item Prefer making the project's REAL build work over disabling it. An env var that
\end{pbitemize}

skips the C extension (e.g. *\_NO\_EXTENSIONS=1) is a last resort, not a shortcut: it runs a different code path from the one the project ships and tests.

\begin{pbitemize}
  \item Do not modify tracked files. Do not run \texttt{git checkout}, \texttt{git clean}, or \texttt{git commit}.
  \item Beware: a dependency that is itself a pytest PLUGIN may be incompatible with the
\end{pbitemize}

version of the project under test at this commit. If the test command starts failing after you install something, remove it.

When the acceptance command works, reply with EXACTLY ONE fenced json block and nothing else -- same fence rules, three backticks and \texttt{json}, no XML tags:

\{output\_spec\}

\pbsub{Task preamble (pytest runner)}

You are configuring a Python project's test environment inside a Linux container.

You are root. The repository is already cloned at \{repo\_dir\} and checked out at commit \{commit\} (\{repo\}). \texttt{uv} and \texttt{git} are installed. You have network access.

\pbsub{Toolchain rules (pytest runner)}

\begin{pbitemize}
  \item Inspect the repo's own declarations first (pyproject.toml [project.optional-dependencies]
\end{pbitemize}

and [dependency-groups], setup.cfg [options.extras\_require], requirements*.txt, tox.ini). Different commits of the same repo declare these DIFFERENTLY.

\begin{pbitemize}
  \item PREFER commands that read the repo at the CURRENT commit -- \texttt{uv pip install -e '.[test]'},
\end{pbitemize}

\texttt{uv pip install -r requirements/test.txt} -- over listing package names by hand. A recipe that reads the repo stays correct at other commits; a hardcoded list does not.

\begin{pbitemize}
  \item Create the venv at \{venv\} with: uv venv --python \{pyver\} \{venv\}
  \item Install with \texttt{uv pip install --python \{venv\}/bin/python ...}.
\end{pbitemize}

\pbsub{Output specification (pytest runner)}

\begin{lstlisting}
["Install": ["<Cmd 1>", "<Cmd 2>"], "Test Cmd": ["<Full-Suite Test Command>"]]
\end{lstlisting}

\texttt{install} must list ONLY the commands that were necessary and succeeded, in order, starting with the \texttt{uv venv} command. \texttt{test\_cmd} is the general (no file selectors) command that runs the whole suite with per-test verbose output and no colour.

NOTE: \texttt{test\_cmd} is PROVENANCE ONLY -- nothing downstream executes it. The grader uses one canonical invocation per test framework, chosen by detecting the framework at this commit, so that a single JUnit parser serves every repo and every language. Your job is the ENVIRONMENT: make the acceptance command above work.

\pbsub{Task preamble (node runner)}

You are configuring a JavaScript/TypeScript project's test environment inside a Linux container.

You are root. The repository is already cloned at \{repo\_dir\} and checked out at commit \{commit\} (\{repo\}). Node \{pyver\}, \texttt{corepack}, \texttt{npm} and \texttt{git} are installed, and you have network access.

USE THE PACKAGE MANAGER THIS COMMIT DECLARES, not the one you would pick. Read package.json's \texttt{packageManager} field first; if it is absent, infer from the committed lockfile -- pnpm-lock.yaml means pnpm, yarn.lock means yarn, package-lock.json means npm, none means npm. Activate it through corepack, e.g. \texttt{corepack prepare pnpm@8.6.3 --activate}. The manager and the lockfile format migrate across this repo's history, so the commit decides, not the repo.

PREFER A FROZEN INSTALL -- \texttt{pnpm install --frozen-lockfile}, \texttt{npm ci}, \texttt{yarn install --immutable}. The committed lockfile pins the whole transitive tree, which is the strongest reproducibility guarantee available here. Fall back to a non-frozen install ONLY if the frozen one fails (an old lockfile can reference a since-yanked package, which no amount of cleverness fixes).

DO NOT add, remove or upgrade a dependency to make the tests run, and do not edit package.json or the lockfile. The JUnit reporters (\texttt{jest-junit}, \texttt{mocha-junit-reporter}) are already installed GLOBALLY and must stay that way -- installing one into the repo would mutate the lockfile and destroy the frozen install.

\pbsub{Toolchain rules (node runner)}

\begin{pbitemize}
  \item Inspect package.json FIRST: \texttt{packageManager}, \texttt{engines}, \texttt{scripts.test}, and which of
\end{pbitemize}

vitest / jest / mocha is in devDependencies. Then check which lockfile is committed. Different commits of the same repo answer these DIFFERENTLY -- the framework and the package manager both migrate across the window.

\begin{pbitemize}
  \item PREFER a lockfile-driven install (\texttt{pnpm install --frozen-lockfile}, \texttt{npm ci},
\end{pbitemize}

\texttt{yarn install --immutable}) over naming packages by hand. It stays correct at other commits; a hardcoded list does not.

\begin{pbitemize}
  \item If the repo is a workspace/monorepo, install at the ROOT. Tests are run per package
\end{pbitemize}

afterwards, but the dependency tree is resolved once at the top.

\begin{pbitemize}
  \item There is NO venv here and nothing to create: \texttt{node\_modules} lands in the repo, which
\end{pbitemize}

is correct and expected. Do not try to install outside the worktree.

\pbsub{Output specification (node runner)}

\begin{lstlisting}
["Install": ["<Cmd 1>", "<Cmd 2>"], "Test Cmd": "<Shell Template, Or Omit>"]
\end{lstlisting}

\pbsub{Task preamble (go runner)}
  
  You are configuring a Go project's test environment inside a Linux container.
  
  You are root. The repository is already cloned at \textbf{[Repo Dir]} and checked out at commit \textbf{[Commit]} (\textbf{[Repo]}). Go \textbf{[Go Version]} and \texttt{git} are installed and you have network access. GOTOOLCHAIN=local is
  set deliberately: this image serves ONE planned Go version, so a go.mod asking for a newer toolchain must fail loudly rather than silently downloading one the plan never chose.
  
  \pbsub{Toolchain rules (go runner)}
  
  \begin{pbitemize}
    \item There is almost never anything to install. \texttt{go test} compiles from source, so the environment is just two warm caches: \texttt{go mod download all} fills \$GOMODCACHE, and a \texttt{go build ./...} fills \$GOCACHE so the 
  first test run does not pay for it. 
    \item If the build fails, the cause is usually a MISSING SYSTEM LIBRARY for a cgo package (sqlite, libgit2, libmagic), not a Go dependency. Install it with apt.
    \item Do NOT edit go.mod or go.sum, and do NOT run \texttt{go mod tidy}. That would change the commit's own dependency graph, which is the thing we are trying to reproduce.
    \item If a package needs a build tag to compile, note it --- but prefer leaving the default build configuration alone.
    \item There is no venv and nothing to create.
  \end{pbitemize}
  
  \pbsub{Output specification (go runner)} 
    
  \begin{lstlisting}
  {"install": ["<cmd 1>", "<cmd 2>"], "test_cmd": "<shell template, or omit>"}
  \end{lstlisting}
    
  \texttt{install} must list ONLY the commands that were necessary and succeeded, in order. For most Go commits that is one or two lines (\texttt{go mod download all}), and an empty list is a legitimate answer when the toolchain alone
  suffices.
  
  The \texttt{test\_cmd} rules are identical to the node runner's (above).

\texttt{install} must list ONLY the commands that were necessary and succeeded, in order.

\texttt{test\_cmd} IS EXECUTED -- it is not documentation. Include it ONLY if running the framework directly on a changed test file does not work. It is a SHELL SNIPPET run once per test file with two variables set:

\$FILE   the test file path, relative to the repo root \$JUNIT  the JUnit XML path it must write

OMIT \texttt{test\_cmd} entirely when the changed file is itself a runnable test -- the default invocation already handles that, and an unnecessary override is a liability.

Include it when the changed file is a FIXTURE: data that some other file turns into a test. Symptom: the run reports 0 tests even though the install is fine. Then find the file that actually contains the test code and write a snippet that derives it from \$FILE, so ONE template works for every chain in this repo. Worked example -- a repo whose tests live in \texttt{\textless{}suite\textgreater{}/test.\textless{}ext\textgreater{}} and whose fixtures sit in \texttt{\textless{}suite\textgreater{}/samples/\textless{}name\textgreater{}/}:

SUITE="\$(dirname "\$(dirname "\$(dirname "\$FILE")")")/test.ts"; \textbackslash{} npx --no-install vitest run "\$SUITE" \textbackslash{} --reporter=junit --outputFile="\$JUNIT"

Derive from \$FILE with dirname/basename -- do NOT hardcode one sample or one suite, or the recipe only works for the chain you happened to test.

DO NOT NARROW THE RUN TO THE CHANGED TEST (no \texttt{-t}, no \texttt{--testNamePattern}, no single sample). Run the whole file or suite that contains it and let us select afterwards. Narrowing looks harmless and quietly destroys the PASS\_TO\_PASS set: the only tests left running are the ones the PR changed, and those FAIL at the pre-fix commit by definition, so nothing is ever observed to pass and the chain ships with no no-regression signal. Measured on svelte -- filtered gave 3 usable outcomes and p2p=0; unfiltered gave 2,995 and cost 19 extra seconds.

\end{promptbox}

\subsection{Problem statement generation}\label{appendix:task_synthesis}

\begin{promptbox}{Task Synthesis Template}
You are a senior engineer writing a pull request description.

\pbhead{Input}

You will receive a code patch (diff) and the original description. Use it only to infer the problem and required behavior. Do NOT describe the implementation.

\pbhead{Goal}

Produce a PR description that explains the problem and what correctness requires, without revealing details unique to this patch.

\pbhead{Rules}

\begin{pbitemize}
  \item Do NOT mention file names, symbols, line numbers, commit hashes, or code snippets.
  \item Do NOT describe algorithms, control flow, or data structures.
  \item Avoid code-level ``changed X to Y'' statements.
  \item High-level component references are allowed.
  \item If intent is unclear, state a brief assumption.
\end{pbitemize}

\pbhead{Format}

\begin{pbitemize}
  \item \textbf{Title} --- one line.
  \item \textbf{Problem} --- 2-3 sentences max.
  \item \textbf{Root Cause} --- 1-2 sentences, conceptual.
  \item \textbf{Fix / Expected Behavior} --- bullet list, max 5 bullets.
  \item \textbf{Risk \& Validation} --- 2-3 bullets total.
\end{pbitemize}

\pbhead{Style Constraints}

\begin{pbitemize}
  \item Prefer bullet points over paragraphs.
  \item No section should exceed 3 lines.
  \item Omit anything non-essential to understanding or reviewing the change.
\end{pbitemize}

\pbhead{Output}

Return only the PR description.

---

\pbhead{Inputs}

\textbf{REPO}: \textbf{[Repo]}

\pbhead{ORIGINAL DESCRIPTION}

\begin{lstlisting}
[Pr Description]
\end{lstlisting}

\pbhead{PATCH**}

\begin{lstlisting}
[Patch]
\end{lstlisting}

\end{promptbox}\label{prompt:task_syn_template}

\begin{promptbox}{Statement Clarity Template}
We have a dataset of GitHub issues from various open-source repositories. Each issue comes with a PR that successfully solves the issue described. Each PR consists of 2 parts: (1) code that resolves the issue (2) changes to the test files of the repository, which check whether the issue has been resolved.

We intend to use samples in this dataset as a benchmark for coding ability: For each sample, we give an engineer the issue text and ask them to write code to resolve the issue (without revealing the solution from the original PR). Then, we apply the test files from the original PR to their code and run the tests to check whether their solution passes.

Importantly, this setup assumes that:

\begin{pbitemize}
  \item The issue description is sufficiently well-specified to understand what the problem is, and what the correct solution should look like.
  \item The tests are correctly scoped to the issue i.e. they correctly test for a solution exactly as described in the issue description, and do not test any other unrelated functionality.
\end{pbitemize}

In this task, you will help to check those assumptions and identify which issue + test samples are suitable for use in our benchmark.

You are now considering an issue from the \textbf{[Repo]} repository.

Please take a moment to read the issue description below.

\textbf{IMPORTANT:} Our setup will only provide the main issue text as shown below, so please answer the questions on the basis of only the text shown here.

\textless{}problem\_statement\textgreater{} \textbf{[Problem Statement]} \textless{}/problem\_statement\textgreater{}

How well-specified is the issue text? Imagine that you are an experienced software engineer who has been instructed to create a PR that successfully resolves the above GitHub issue. You have full access to the codebase, and can see the issue description as it is above. But you are not able to ask for clarification and would need to work exclusively from this information. Is the issue description well-specified enough for a meaningful attempt at a solution?

You should classify this issue based on the following options.

\begin{pbitemize}
  \item 0: The issue is well-specified and it is clear what is required for a successful solution.
\end{pbitemize}

  \textless{}example\textgreater{} Title: \texttt{parse\_date} raises on ISO strings that carry a timezone offset. Problem: Calling \texttt{parse\_date('2021-01-01T00:00:00+05:00')} raises \texttt{ValueError: unconverted data remains:
  +05:00}. It should instead return a timezone-aware datetime carrying the given UTC offset. Expected behavior: \texttt{parse\_date('2021-01-01T00:00:00+05:00')} returns \texttt{datetime(2021, 1, 1, 0, 0,
  tzinfo=timezone(timedelta(hours=5)))}. Naive inputs such as \texttt{parse\_date('2021-01-01')} already work and must keep returning a naive datetime unchanged. \textless{}/example\textgreater{}

\begin{pbitemize}
  \item 1: There are some blanks to fill in about the issue, but there is a sensible interpretation of what is required for a successful solution.
\end{pbitemize}

  \textless{}example\textgreater{} Title: HTTP client should not retry on 4xx responses. The client currently retries every failed request up to its configured maximum. Client errors (4xx) will not succeed on retry, so they should be
  returned immediately instead of retried; only network errors and 5xx responses should be retried. Existing retry count and backoff behavior for the retryable cases should stay as they are. \textless{}/example\textgreater{}

\begin{pbitemize}
  \item 2: The issue is vague and there is room for ambiguity. It is unclear what a successful solution would look like.
\end{pbitemize}

  \textless{}example\textgreater{} Title: Improve error messages for config parsing. The error messages produced when a configuration file is malformed are not very helpful. They should be made clearer so the user can understand what went
  wrong and fix their config. \textless{}/example\textgreater{}

\begin{pbitemize}
  \item 3: It is almost impossible to understand what you are being asked to do without further information.
\end{pbitemize}

  \textless{}example\textgreater{} Title: Make the scheduler more robust. The scheduler sometimes behaves unexpectedly when the system is under load. We should make it more robust and reliable. See the discussion in the linked thread for the
  background and context. \textless{}/example\textgreater{}

Please respond with a short explanation of your choice (minimum 100 characters) followed by the selected option number.\textless{}/pre\textgreater{}

\end{promptbox}

  \begin{promptbox}{Statement Leakage Template}
  You are auditing an agent-facing PROBLEM STATEMENT to ensure it is a FAIR test of whether an agent ANTICIPATES an edge case.
  
  Setup: an earlier PR (PR1) introduced/changed behavior in \textbf{[Repo]}. A later PR (PR2) then fixed an edge case PR1 missed. We give an agent PR1's problem statement, have it solve PR1 from clean intent, and score whether its solution
  passes PR2's held-out tests --- i.e.\ whether it ANTICIPATED the edge case WITHOUT being told.
  
  For this to be fair the statement must sit in a Goldilocks zone:
  
  \begin{pbitemize}
    \item state the GOAL so a competent solution is possible;
    \item NOT prescribe the specific (flawed) implementation PR1 used --- else following the task faithfully FORCES the bug, so the task contradicts the anticipation (UNFAIR);
    \item NOT reveal the edge-case handling PR2's tests check --- else the agent is simply told the answer (LEAKY, anticipation becomes trivial).
  \end{pbitemize}
  
  \pbhead{The edge case --- ANSWER KEY (the agent must NOT be able to derive this from the statement)}
  
  PR2's fix (what the missed edge case was): \textless{}\textless{}\textless{} \textbf{[Anticipation Patch]} \textgreater{}\textgreater{}\textgreater{}
  
  PR2's held-out tests (what ``anticipation'' is scored on): \textless{}\textless{}\textless{} \textbf{[Anticipation Test Patch]} \textgreater{}\textgreater{}\textgreater{}
  
  \pbhead{PR1's actual change (the introducer --- may embody the flawed mechanism)}
  
  \textless{}\textless{}\textless{} \textbf{[Pr1 Patch]} \textgreater{}\textgreater{}\textgreater{}
  
  \pbhead{The PROBLEM STATEMENT under audit (what the agent sees)}
  
  \textless{}\textless{}\textless{} \textbf{[Pr1 Statement]} \textgreater{}\textgreater{}\textgreater{}
  
  \pbhead{Decide the PRIMARY verdict}
  
  \begin{pbitemize}
    \item "clean" --- intent-level: a correct anticipating solution satisfies it, and it neither prescribes the flawed mechanism nor reveals the edge-case handling.
    \item "prescribes\_mechanism" --- dictates the specific implementation/approach that produces the missed edge case (a control-flow or data-structure choice PR2 had to undo), so faithfully following it forces the flaw.
    \item "leaks\_anticipation" --- states or clearly implies the CORRECT handling of the edge case PR2's tests verify (the thing the agent should have to DISCOVER on its own).
    \item "leaks\_future\_fix" --- references the later fix / PR2 / the bug report / the specific failure symptom the held-out tests target.
  \end{pbitemize}
  
  Severity order if several apply: prescribes\_mechanism \textgreater{} leaks\_future\_fix \textgreater{} leaks\_anticipation \textgreater{} clean.
  
  Be strict about "leaks\_anticipation": any sentence that tells the agent how the edge-case-relevant cases must behave (e.g.\ ``multi-result methods must still collect all matches'', ``empty input returns X'', ``must guard against null'')
  is a leak --- the agent should derive that.
  
  \pbhead{Output (JSON only)}
  \begin{lstlisting}
  {
    "verdict": "clean|prescribes_mechanism|leaks_anticipation|leaks_future_fix",
    "evidence": "the exact phrase from the statement that triggers it, or 'none'",
    "reasoning": "2-3 sentences tying the statement to the specific edge case",
    "repair_hint": "how to rewrite to intent-only: what to remove, what to keep"
  } 
  \end{lstlisting}
  \end{promptbox}

\subsection{Dataset validation}

More than 90\% of our corpus relies on SZZ to identify $\text{PR}_1$. Although SZZ is widely used to identify bug-introducing changes, its line-level blame heuristic can produce incorrect attributions~\citep{lyu2026agentszzteachingllmagent, cao2026masszzmultiagenticszzalgorithm, 10.1145/3728885}. For example, a later refactoring may overwrite the original blame, causing SZZ to identify the refactor rather than the change that introduced the defect. A separate failure mode occurs when the $\text{PR}_2$ fix is implemented in a different file or function from the change that originally caused the defect. Because SZZ traces the history of lines modified by $\text{PR}_2$, it relies on overlap between the fix location and the causal change; when this overlap is absent, the inferred $\text{PR}_1$ can be incorrect. SZZ can only trace blame through lines that are modified or deleted by $\text{PR}_2$. If the fix consists entirely of newly added lines, there is no pre-existing line to blame back to an earlier change, so SZZ cannot infer a $\text{PR}_1$ and abstains. 

To reduce contamination from incorrectly inferred $\text{PR}_1$s, we use a panel of LLM judges to filter candidate pairs before they enter the benchmark. Specifically, we use DeepSeek V4 Flash~\citep{deepseekv4} and GLM 5.3 Flash~\citep{glm53}. We further evaluate the reliability of this filtering setup against ground-truth $\text{PR}_1$s explicitly identified by maintainers in PR and issue discussions. We estimate a contamination rate of 7.3\% (95\% CI [4.2, 12.4]). Importantly, this estimate should be interpreted as a strict upper bound on contamination in the final benchmark: of the 252 SZZ-derived candidates considered in this analysis, only 28 survive the subsequent stages of our data-curation pipeline.

\textbf{Dataset.} A subset of fix chains includes an explicit human attribution identifying $\text{PR}_1$, typically stated by a maintainer in the associated PR or issue discussion. Our dataset contains 376 instances across 46 repositories and 4 programming languages, representing 2.8\% of all mined instances. For these chains the true $\text{PR}_1$ is known independently of SZZ, which lets us measure both how often SZZ's inferred $\text{PR}_1$ is correct and how much of its error the judges remove.

\textbf{Results.} We run the SZZ algorithm on the 376 instances in our dataset. Of 376 chains it abstains on 114 and proposes a $\text{PR}_1$ for 262, of which 172 match the maintainer and 90 do not: a contamination rate of 34.4\% before running LLM-judge panel. The LLM-judge panel rejects 85.6\% of the incorrect $\text{PR}_1$s (77 of 90) while discarding 10.5\% of the correct ones (18 of 172). Contamination among retained chains falls from 34.4\% to 7.8\% (95\% CI [4.6, 12.9]), at a yield of 89.5\%

Further, validate the entire Foresight-SWE construction pipeline end-to-end by sampling 10\% of all instances and ask a co-author who was not involved in the benchmark construction process whether each instance has a PR$_2$ that fixes a bug introduced by PR$_1$.
The co-author found that 90.9\% of validated instances contained such PR$_2$s and the 9\% that didn't were debatably so.

\section{Dataset Annotation}\label{appendix:data_annotation}

\subsection{Instance difficulty and severity annotation}\label{appendix:difficulty_severity}

We use three prompts in sequence to produce the reported difficulty and defect severity. The first identifies the functions and classes exercised by the anticipation tests. The second prompt assigns a difficulty label, based on estimated engineering effort, and task validity code. The difficulty definitions are given in Table~\ref{tab:categories_def}. Both prompts are taken verbatim from SWE-ReBench V2 \citep{swerebenchv2}. The third prompt is our own and assigns the severity of the defect. The severity definitions of the defect are given in Table~\ref{tab:categories_def}. Inspired by ODC~\citep{odc} we define three axis and based on their values annotate an instance as critical, major or minor. The first axis is impact which measures what is affected by the defect. The second axis is reach and it measures how common the triggering setup is among it's users and according to the repositories documentation. Lastly, detectability which measures whether the user sees an error or silently receives a wrong result, inferred from the held-out test's assertions. A fixed rule maps these facts to a tier (Table~\ref{tab:severity_rule}). We use DeepSeek-V4 Flash~\citep{deepseekv4} for all three prompts.

We use DeepSeek-V4 Flash~\cite{deepseekv4} to judge task validity and assign difficulty and defect-severity labels.

\begin{table}[H]
  \centering
  \small
  \caption{\textbf{Category definitions.} The two axes along which we stratify the
  benchmark. \emph{Difficulty} is scored by an LLM judge from the amount of code change, logic complexity, and domain knowledge required, expressed as expected implementation
  time, and is applied to both the original task (PR$_1$) and the fix (PR$_2$). \emph{Severity} are judged on PR$_2$ (the fix).}
  \renewcommand{\arraystretch}{1.25}
  \begin{tabularx}{\textwidth}{@{}l l X@{}}
    \toprule
    \textbf{Axis} & \textbf{Category} & \textbf{Definition} \\
    \midrule
    \multirow{3}{*}{\shortstack[l]{\textbf{Difficulty}\\[2pt]\footnotesize (PR$_1$ \& PR$_2$)}}
      & Easy   & Expected developer implementation time $<15$ min. \\
      & Medium & Expected implementation time $15$ min--$1$h. \\
      & Hard   & Expected implementation time $>1$h. \\
    \midrule
    \multirow{3}{*}{\textbf{Severity}}
        & Critical & Persisted data corruption, data loss, or a security bypass reachable from a default or mainstream setup. \\
        & Major    & Broken or wrong output on a mainstream path; or silent wrong output on a niche path; or corruption behind a niche trigger. \\
        & Minor    & Loud failure on a niche path; any failure needing a rare intersection of conditions; or degraded/cosmetic impact. \\
    \bottomrule \\
  \end{tabularx}
  \label{tab:categories_def}
\end{table}

  \begin{table}[H]
    \centering
    \small
    \caption{\textbf{Severity rule.} The judge extracts impact, reach and detectability and the tier is computed from them. }
    \begin{tabular}{@{}lccc@{}}
      \toprule
      \textbf{Impact} $\backslash$ \textbf{Reach} & \textbf{Broad} & \textbf{Narrow} & \textbf{Rare} \\
      \midrule
      Corruption          & Critical & Major & Major \\
      Broken              & Major    & Major if silent, else Minor & Minor \\
      Degraded / Cosmetic & Minor    & Minor & Minor \\ 
      \bottomrule
    \end{tabular}         
    \label{tab:severity_rule} 
  \end{table}

\begin{promptbox}{Interface Extraction Template}
You analyze code changes in pull requests. Your task is to extract information about changed interfaces and signatures that are explicitly checked in the tests.

\pbhead{Decision Process}

\begin{pbenumerate}
  \item Identify new functions or classes
  \item Identify functions or classes whose signatures have changed significantly (arguments changed, not just indentation)
  \item Use the test patch to see which of them, if any, are explicitly called in the new test code
\end{pbenumerate}

Ignore as interfaces:

\begin{pbitemize}
  \item Purely internal helpers (for example names starting with "\_" and not used directly in tests)
  \item Type aliases, enums, simple data or config holders
  \item Constructors or trivial overrides that only support other documented interfaces
  \item Constants, flags, small config values, and simple data declarations
  \item Anything that is not explicitly called in the test patch
\end{pbitemize}

\pbhead{Output Format}

Output must be wrapped in a single root tag:

\begin{lstlisting}
<ANSWER>Method: <ClassName.method_name(self, key_params...)> Location:  Inputs: <important parameters, types or roles, and key constraints> Outputs: <return type or behavior, including important error conditions> Description: <1 to 2 sentences on what it does and when to use it>

Function: <function_name(key_params...)>
Location:
Inputs: <important parameters, types or roles, and key constraints>
Outputs: <return type or behavior, including important error conditions>
Description: <1 to 2 sentences on what it does and when to use it>

...repeat blocks for each interface...</ANSWER>
\end{lstlisting}

If there are no qualifying interfaces, output exactly:

\begin{lstlisting}
<ANSWER>No new interfaces are introduced.</ANSWER>
\end{lstlisting}

Guidelines:

\begin{pbitemize}
  \item Only include symbols that:
  \item appear in the production patch with a changed signature as new or clearly changed, and
  \item are directly constructed or called in \textbf{[Test Patch]}
  \item Keep descriptions technical and concise
\end{pbitemize}

---

\pbhead{Inputs}

\textbf{REPO}: \textbf{[Repo]}

\pbhead{ISSUE\_TEXT**}

\begin{lstlisting}
[Problem Statement]
\end{lstlisting}

\pbhead{TEST\_PATCH**}

\begin{lstlisting}
[Test Patch]
\end{lstlisting}

\textbf{GOLDEN PATCH} (reference only):

\begin{lstlisting}
[Patch]
\end{lstlisting}

\end{promptbox}

\begin{promptbox}{Task Validity and Difficulty}
You are classifying software engineering tasks for RL validity. Determine if a task is of high-quality or reflects environment preparation issues (B1-B11).

\pbhead{Classification Codes}

{\footnotesize
\begin{tabularx}{\linewidth}{@{}llX@{}}
\toprule
  \textbf{Code} & \textbf{Category} & \textbf{Description} \\
\midrule
  A & SOLVABLE & Problem is clearly specified, tests align with stated requirements \\
  B1 & TEST\_SUITE\_COUPLING & Correct fix may fail due to unrelated test regressions \\
  B2 & IMPLICIT\_NAMING & Tests expect specific names, signatures, error message strings, or output values derived from the patch implementation but not specified in the problem statement \\
  B3 & EXTERNAL\_DEPENDENCY & Essential info in external URLs not quoted in issue \\
  B4 & AMBIGUOUS\_SPEC & Missing expected behavior, repro steps, or acceptance criteria \\
  B5 & PATCH\_ARTIFACTS & Reference patch includes unrelated changes tests may require \\
  B6 & IMPLICIT\_KNOWLEDGE & Requires domain knowledge not stated or inferrable from repo \\
  B7 & INLINE\_TEST & Inline test in the patch \\
  B8 & VISUAL\_REQUIREMENT & Task requires images/visualizations/visual UI context not captured in plain text --- including inline markdown images (\texttt{![...](url)}) embedded in the issue or PR description \\
  B9 & NON\_ENGLISH\_TEXT & Task description contains non-English language that must be understood \\
  B10 & TEXT\_FILE\_TESTS & Test patch only modifies plain-text fixture files (e.g. \texttt{.txt}, \texttt{.csv}) rather than code; expected behavior is encoded as opaque text diffs, making it hard to understand what is being tested \\
  B11 & MULTI\_ISSUE\_PR & PR addresses multiple independent issues or bugs simultaneously, creating ambiguity about which change is required to pass the tests \\
\bottomrule
\end{tabularx}}

\pbhead{Decision Process}

\begin{pbenumerate}
  \item Extract intent from ISSUE\_TEXT: expected behavior, observed behavior, reproduction, acceptance criteria
  \item Check test alignment: Do assertions verify stated requirements or introduce new ones?
  \item Scan for B-category signals:
\end{pbenumerate}

\begin{pbitemize}
  \item B1: Tests span unrelated modules
  \item B2: String assertions for names, error messages, or specific output values not derivable from the issue; assertIn/assertEqual on literal strings present only in the golden patch
  \item B3: URLs to external docs/specs
  \item B4: Vague or missing acceptance criteria
  \item B5: Patch modifies unrelated files that are then required by tests
  \item B6: Tests assert specific approaches not mentioned
  \item B7: Inline test in the normal PATCH
  \item B8: Visual artifacts are required to understand/solve task --- check for markdown image syntax \texttt{![...](url)} in ISSUE\_TEXT, or references to screenshots/mockups that cannot be read as plain text
  \item B9: Issue/problem statement is partially or fully non-English
  \item B10: All changed files in TEST\_PATCH are \texttt{.txt}, \texttt{.csv}, or similar non-code fixture files; no Python/JS/etc. test functions are added or modified
  \item B11: Problem statement or PR references/closes multiple separate GitHub issues; tests appear to cover two or more independent fixes
\end{pbitemize}

\begin{pbenumerate}
  \item Classify: A if clean, B[1-11] for primary issue
\end{pbenumerate}

\pbhead{Rubric: PR Category}

Choose one or more categories that best describe the PR.

Allowed categories:

\begin{lstlisting}
{
  "critical_bug", "major_bug", "minor_bug", "regression_bug", "edge_case_bug", "performance_bug",
  "security_bug", "integration_feat", "core_feat", "ui_ux_feat", "dev_ops_enh", "documentation_enh"
}
\end{lstlisting}

\pbhead{Rubric: Task Difficulty}

Estimate task difficulty based on the amount of code changes required, logic complexity, and domain knowledge needed. Choose one level assuming expected time to implement: easy (\textless{}15 min), medium (15 min--1h), hard (\textgreater{}1h).

Allowed difficulty levels:

\begin{lstlisting}
{
  "easy", "medium", "hard"
}
\end{lstlisting}

\pbhead{Output (JSON only)}

\begin{lstlisting}
{
  "reasoning": "3-5 sentences: (1) summarize issue intent, (2) assess test alignment, (3) note any B-category signals, (4) justify classification",
  "intent_completeness": "complete | partial | insufficient",
  "test_alignment_issues": ["list specific misalignments or empty if aligned"],
  "detected_issues": {
    "B1": false, "B2": false, "B3": false,
    "B4": false, "B5": false, "B6": false, "B7": false,
    "B8": false, "B9": false, "B10": false, "B11": false
  },
  "external_urls": ["extracted URLs if any"],
  "pr_categories": ["one or more from the PR category rubric"],
  "difficulty": "easy | medium | hard",
  "code": "A|B1|B2|B3|B4|B5|B6|B7|B8|B9|B10|B11",
  "confidence": 0.0
}

\end{lstlisting}

---

\pbhead{Inputs}

\textbf{REPO}: \textbf{[Repo]}

\pbhead{ISSUE\_TEXT**}

\begin{lstlisting}
[Problem Statement]

These are interfaces that you need to implement, they will be tested:
[Interface]
\end{lstlisting}

\pbhead{TEST\_PATCH**}

\begin{lstlisting}
[Test Patch]
\end{lstlisting}

\textbf{GOLDEN PATCH} (reference only):

\begin{lstlisting}
[Patch]
\end{lstlisting}

\end{promptbox}

  \begin{promptbox}{Defect Severity Template}
  You are a senior software maintainer characterizing a real bug for a research dataset. An earlier pull request (PR1) in \textbf{[Repo]} introduced or changed some behavior and missed an edge case; a later PR (PR2) fixed that missed edge
  case. Your job is to EXTRACT STRUCTURED FACTS about the bug PR2 fixed --- one value per axis, each backed by a short quote from the evidence.
  
  Do NOT output a severity tier. Severity is COMPUTED from your axis values by a downstream function. Your job is to get the axes right, not to grade. Judge from the perspective of a USER of the library, between PR1 and PR2. This judge runs
  only on chains already confirmed to be genuine bugs, so every input IS a real user-facing bug.
  
  \pbhead{Evidence}
  
  Repository: \textbf{[Repo]}
  
  PR1's task (what the earlier change set out to do): \textless{}\textless{}\textless{} \textbf{[Pr1 Task]} \textgreater{}\textgreater{}\textgreater{}
  
  PR2's description (human-written fix for the missed edge case): \textless{}\textless{}\textless{} \textbf{[Pr2 Description]} \textgreater{}\textgreater{}\textgreater{}
  
  PR2's fix patch: \textless{}\textless{}\textless{} \textbf{[Fix Patch]} \textgreater{}\textgreater{}\textgreater{}
  
  PR2's test patch (the held-out tests --- fail before the fix, pass after; read the assertions to reason about the pre-fix behavior): \textless{}\textless{}\textless{} \textbf{[Test Patch]} \textgreater{}\textgreater{}\textgreater{}
  
  Human discussion (PR + linked issues/tickets; each item carries a URL): \textless{}\textless{}\textless{} \textbf{[Discussion]} \textgreater{}\textgreater{}\textgreater{}
  
  Objective evidence (gap, held-out test ids, changelog lines): \textless{}\textless{}\textless{} \textbf{[Evidence]} \textgreater{}\textgreater{}\textgreater{}
  
  \pbhead{1. impact --- the CONSEQUENCE to a user (not the mechanism)}
  
  \begin{pbitemize} 
    \item "corruption" --- PERSISTED or irreversible wrong data (written to a DB / file / durable state), data loss, OR a security/authorization bypass. The dangerous, hard-to-undo class; the user trusts a wrong answer that sticks.
    \item "broken" --- a documented capability produces WRONG or MISSING output, or stops working: a wrong value / URL / string; a false-positive diagnostic (the feature reports a wrong verdict); an exception where a result was owed; a
  feature so defeated it is unusable for its purpose (e.g.\ resume that forces a full re-download); a control-flow contract silently altered; OR a standard workflow halts (e.g.\ warning spam failing a zero-warning CI gate). Wrong but
  TRANSIENT / recoverable output is "broken", NOT "corruption".
    \item "degraded" --- still produces CORRECT results and blocks no workflow, but worse: perf regression, non-fatal noise.
    \item "cosmetic" --- output formatting, messages, listings.
  \end{pbitemize}
    
  Judge from the DOWNSTREAM consumer, assuming the least-informed one.
    
  \pbhead{2. detectability --- does the user OBSERVE it?}
    
  Reason from the held-out test's assertions about what the PRE-FIX behavior was:
    
  \begin{pbitemize}
    \item "silent" --- pre-fix returns a WRONG VALUE (assertion on a value), or an EXPECTED EXCEPTION IS MISSING (the fix adds a raise / the error was swallowed).
    \item "loud" --- pre-fix raises an UNEXPECTED exception / crash the user sees.
  \end{pbitemize}
    
  \pbhead{3. reach --- how COMMON is the setup needed to trigger it?}
    
  Decompose the trigger into its conditions (from the test's arrange/act), then classify by the RAREST (limiting) condition --- NOT by ``is there a flag''.
  \begin{pbitemize}
    \item "broad" --- default settings, OR a mainstream usage pattern.
    \item "narrow" --- a specific niche opt-in (obscure config/dialect/platform, or a minority API).
    \item "rare" --- an intersection of $\geq$2 unusual conditions.
  \end{pbitemize}
  
  CRITICAL: opt-in $\neq$ narrow. A single opt-in to a COMMON setting is broad. (E.g.\ requiring \texttt{-{}-import-mode=importlib} is NOT narrow --- pytest recommends it as the future default; the limiting condition is whatever ELSE the
  test needs.) You MUST justify prevalence with cited evidence: a documented default/recommendation, the library's own tutorial usage, OR the count of independent reporters in the discussion. No evidence -\textgreater{} output "narrow" and
  set "reach\_estimated": true. 
  
  \pbhead{4. security --- an ORTHOGONAL yes/no tag (a bug can be security AND have any severity)}
  
  "is\_security": true/false --- does the fix close an attack (injection, XSS, path traversal, auth/authorization bypass, CRLF/header injection, weak randomness, ReDoS/DoS, SSRF, info leak)? NOTE: this is a tag ONLY --- it does NOT change
  the severity. Judge impact/reach/detectability on the actual consequence regardless (a defense-in-depth hardening fix with no demonstrated exploit can still be low impact).
  
  \pbhead{5. priority\_evidence --- REALIZED signals (record; they DO NOT set severity)}
  
  Version pins/downgrades, reaction counts, number of independent reporters, a maintainer hotfix release. These are urgency (IEEE-1044 Priority), used only to corroborate the reach estimate, never the tier. Quote them; mark realized=true.
  
  \pbhead{6. mechanism --- did PR1 break working behavior, or miss a case in new code?}
  
  \begin{pbitemize}
    \item "regression" --- something that worked before PR1 stopped (PR1 broke it).
    \item "incomplete\_coverage" --- PR1's new behavior was correct but missed a case in its own domain.
  \end{pbitemize}
  
  \pbhead{Output (JSON only, NO tier)}
  
  \begin{lstlisting}
  {
    "impact":        {"value": "corruption|broken|degraded|cosmetic", "evidence": "short quote"},
    "detectability": {"value": "silent|loud", "evidence": "the assertion you reasoned from"},
    "reach":         {"value": "broad|narrow|rare", "limiting_condition": "the rarest condition",
                      "prevalence_evidence": "doc default / tutorial / N reporters / or 'none'",
                      "reach_estimated": false},
    "security":      {"is_security": false, "evidence": ""},
    "priority_evidence": {"signals": "pins/reactions/reporters/hotfix or 'none'", "realized": true},
    "mechanism":     {"value": "regression|incomplete_coverage", "evidence": "short quote"},
    "user_impact":   "1-2 sentences: what a user of the library experiences",
    "confidence":    0.0
  } 
  \end{lstlisting}
  \end{promptbox}

\section{Additional Experimental Setup Details}\label{appendix:exp_setup_detail}

\begin{table}[H]
      \centering
      \small
      \begin{tabular}{l l l}
        \toprule
        \textbf{Agent Harness} & \textbf{Version} & \textbf{Reference} \\
        \midrule
        mini-swe-agent  & 1.9.1   & \url{https://github.com/swe-agent/mini-swe-agent} \\
        Claude Code CLI & 2.1.278 & \url{https://claude.com/product/claude-code} \\
        Codex CLI       & 0.155.0 & \url{https://openai.com/codex/} \\
        \bottomrule
      \end{tabular} 
      \caption{Agent harnesses used in this work.}
      \label{tab:harness_versions}
  \end{table}
      
  \begin{table}[H]
      \centering
      \small
      \begin{tabular}{l c c c}
        \toprule
        \textbf{Model} & \textbf{Steps / Turns} & \textbf{Cost limit} & \textbf{Timeout} \\
        \midrule
        GLM-5.3 Flash         & 250 & \$3    & --- \\
        Qwen3.8 Flash Next    & 250 & \$3    & --- \\
        Qwen3.5-397B-A17B     & 250 & \$3    & --- \\
        DeepSeek-V4 Flash     & 350 & \$3    & --- \\
        Qwen3.8-27B           & 250 & \$3    & --- \\
        Qwen3.5-27B           & 250 & \$3    & --- \\
        \midrule
        Claude Opus 5         & 250 & \$5    & 90 min \\
        Claude Sonnet 5       & 250 & \$5    & 90 min \\
        \midrule
        GPT-5.6 Sol           & --- & ---    & 90 min \\
        GPT-5.6 Terra         & --- & ---    & 90 min \\
        GPT-5.6 Luna          & --- & ---    & 90 min \\
        \bottomrule           
      \end{tabular}
      \caption{\textbf{Per model budgets.} Steps apply to mini-swe-agent and turns to Claude Code. Codex exposes neither a turn nor a cost limit, so it is bounded only by the wall-clock timeout. All models use reasoning effort as high except Qwen3.8 Flash Next which has only xhigh.}
      \label{tab:model_budgets}
  \end{table}

\subsection{Proactive Prompt Template}
The below prompt is attached at the end of the problem statement for each harness.

\begin{promptbox}{Proactive Prompt}
  \textless{}important\textgreater{}

  The PR description above states the behavior that was asked for. It covers the intent of the request, but it may not cover every edge case that behavior has to handle. You should attempt to forecast any potential edge cases that your
  implementation may not cover, or edge cases introduced by your implementation.
  
  Work those out yourself, from the codebase: read how the code you are changing is called, what its neighbours already handle, and which inputs and expected outputs the description never names. Do this before you start, while you implement,
  and again before you submit --- what you learn from the code as you go will change what counts as an edge case here.
  
  One limit. Stay inside the scope of the requested behavior: cover the inputs of THAT behavior, and do not redesign or harden unrelated parts of the codebase.
  
  Verify the cases you decide to handle with your own scratch script, not by changing the repository's tests.
  
  \textless{}/important\textgreater{}
  \end{promptbox}

\section{Additional Results}
\subsection{Stratified results by task difficulty, defect severity and defect detectability}\label{appendix:strati_full_result}
\begin{table}[H]
    \centering
    \caption{\textbf{Per-model results stratified by task difficulty.} Resolve rate and FRR for all eleven models.}
    \begin{tabular}{lrrrrrr}
        \toprule
        & \multicolumn{3}{c}{$\text{PR}_1$ difficulty} & \multicolumn{3}{c}{$\text{PR}_2$ difficulty} \\
        \cmidrule(lr){2-4}\cmidrule(lr){5-7}
        \textbf{Model} & \textbf{Easy} & \textbf{Medium} & \textbf{Hard} & \textbf{Easy} & \textbf{Medium} & \textbf{Hard} \\
        & {\footnotesize $n$=47} & {\footnotesize $n$=77} & {\footnotesize $n$=88} & {\footnotesize $n$=125} & {\footnotesize $n$=52} & {\footnotesize $n$=35} \\
        \midrule
        \multicolumn{7}{l}{\emph{Resolve rate}} \\ \midrule
        \quad Qwen3.8 & 95.7 & 97.4 & 85.2 & 91.2 & 98.1 & 85.7 \\
        \quad GLM-5.3 & 76.6 & 80.5 & 83.0 & 73.6 & 90.4 & 91.4 \\
        \quad DeepSeek-V4 & 68.1 & 59.7 & 52.3 & 55.2 & 69.2 & 54.3 \\
        \quad GPT-5.6 Sol & 93.6 & 81.8 & 81.8 & 80.0 & 90.4 & 91.4 \\
        \quad GPT-5.6 Terra & 85.1 & 79.2 & 64.8 & 76.8 & 71.2 & 71.4 \\
        \quad GPT-5.6 Luna & 87.2 & 76.6 & 64.8 & 72.0 & 76.9 & 77.1 \\
        \quad Opus 5 & 85.1 & 74.0 & 77.3 & 74.4 & 88.5 & 74.3 \\
        \quad Sonnet 5 & 76.6 & 57.1 & 53.4 & 56.8 & 65.4 & 62.9 \\
        \quad Qwen3.5-397B & 46.8 & 23.4 & 6.8 & 22.4 & 23.1 & 17.1 \\
        \quad Qwen3.8-27B & 53.2 & 27.3 & 9.1 & 25.6 & 28.8 & 20.0 \\
        \quad Qwen3.5-27B & 38.3 & 23.4 & 3.4 & 17.6 & 25.0 & 11.4 \\
        \midrule
        \multicolumn{7}{l}{\emph{FRR}} \\ \midrule
        \quad Qwen3.8 & 27.7 & 31.2 & 8.0 & 28.8 & 15.4 & 0.0 \\
        \quad GLM-5.3 & 17.0 & 14.3 & 5.7 & 12.0 & 15.4 & 2.9 \\
        \quad DeepSeek-V4 & 21.3 & 29.9 & 8.0 & 21.6 & 23.1 & 2.9 \\
        \quad GPT-5.6 Sol & 21.3 & 23.4 & 12.5 & 24.8 & 15.4 & 0.0 \\
        \quad GPT-5.6 Terra & 25.5 & 16.9 & 8.0 & 20.8 & 9.6 & 2.9 \\
        \quad GPT-5.6 Luna & 23.4 & 23.4 & 6.8 & 20.8 & 17.3 & 0.0 \\
        \quad Opus 5 & 38.3 & 26.0 & 15.9 & 28.0 & 30.8 & 2.9 \\
        \quad Sonnet 5 & 21.3 & 20.8 & 11.4 & 19.2 & 19.2 & 5.7 \\ 
        \quad Qwen3.5-397B & 10.6 & 11.7 & 1.1 & 8.8 & 5.8 & 2.9 \\
        \quad Qwen3.8-27B & 12.8 & 10.4 & 2.3 & 8.8 & 7.7 & 2.9 \\
        \quad Qwen3.5-27B & 10.6 & 9.1 & 1.1 & 7.2 & 7.7 & 0.0 \\
        \bottomrule
    \end{tabular}
\end{table}

\begin{table}[H]
    \centering
    \caption{\textbf{Per-model results stratified by severity and detectability.} Resolve rate and FRR for all eleven models.}
    
     \begin{tabular}{lrrrr}
       \toprule
       & \multicolumn{2}{c}{Severity} & \multicolumn{2}{c}{Detectability} \\
       \cmidrule(lr){2-3}\cmidrule(lr){4-5}
       \textbf{Model} & \textbf{Minor} & \textbf{Major} & \textbf{Loud} & \textbf{Silent} \\
       & {\footnotesize $n$=83} & {\footnotesize $n$=129} & {\footnotesize $n$=83} & {\footnotesize $n$=129} \\
       \midrule
       \multicolumn{5}{l}{\emph{Resolve rate}} \\ \midrule
       \quad Qwen3.8 & 91.6 & 92.2 & 92.8 & 91.5 \\
       \quad GLM-5.3 & 79.5 & 81.4 & 80.7 & 80.6 \\
       \quad DeepSeek-V4 & 59.0 & 58.1 & 61.4 & 56.6 \\
       \quad GPT-5.6 Sol & 88.0 & 82.2 & 90.4 & 80.6 \\
       \quad GPT-5.6 Terra & 77.1 & 72.9 & 83.1 & 69.0 \\
       \quad GPT-5.6 Luna & 74.7 & 73.6 & 78.3 & 71.3 \\
       \quad Opus 5 & 77.1 & 78.3 & 78.3 & 77.5 \\
       \quad Sonnet 5 & 62.7 & 58.1 & 65.1 & 56.6 \\
       \quad Qwen3.5-397B & 24.1 & 20.2 & 24.1 & 20.2 \\
       \quad Qwen3.8-27B & 25.3 & 25.6 & 26.5 & 24.8 \\
       \quad Qwen3.5-27B & 18.1 & 18.6 & 16.9 & 19.4 \\
       \midrule
       \multicolumn{5}{l}{\emph{FRR}} \\ \midrule
       \quad Qwen3.8 & 20.5 & 20.9 & 25.3 & 17.8 \\
       \quad GLM-5.3 & 13.3 & 10.1 & 15.7 & 8.5 \\
       \quad DeepSeek-V4 & 24.1 & 15.5 & 24.1 & 15.5 \\
       \quad GPT-5.6 Sol & 18.1 & 18.6 & 20.5 & 17.1 \\
       \quad GPT-5.6 Terra & 18.1 & 13.2 & 18.1 & 13.2 \\ 
       \quad GPT-5.6 Luna & 18.1 & 15.5 & 20.5 & 14.0 \\
       \quad Opus 5 & 33.7 & 18.6 & 32.5 & 19.4 \\
       \quad Sonnet 5 & 22.9 & 13.2 & 21.7 & 14.0 \\ 
       \quad Qwen3.5-397B & 10.8 & 4.7 & 9.6 & 5.4 \\
       \quad Qwen3.8-27B & 9.6 & 6.2 & 10.8 & 5.4 \\
       \quad Qwen3.5-27B & 7.2 & 5.4 & 4.8 & 7.0 \\
       \bottomrule
     \end{tabular}
\end{table}

\begin{figure}
    \centering
    \includegraphics[width=\linewidth]{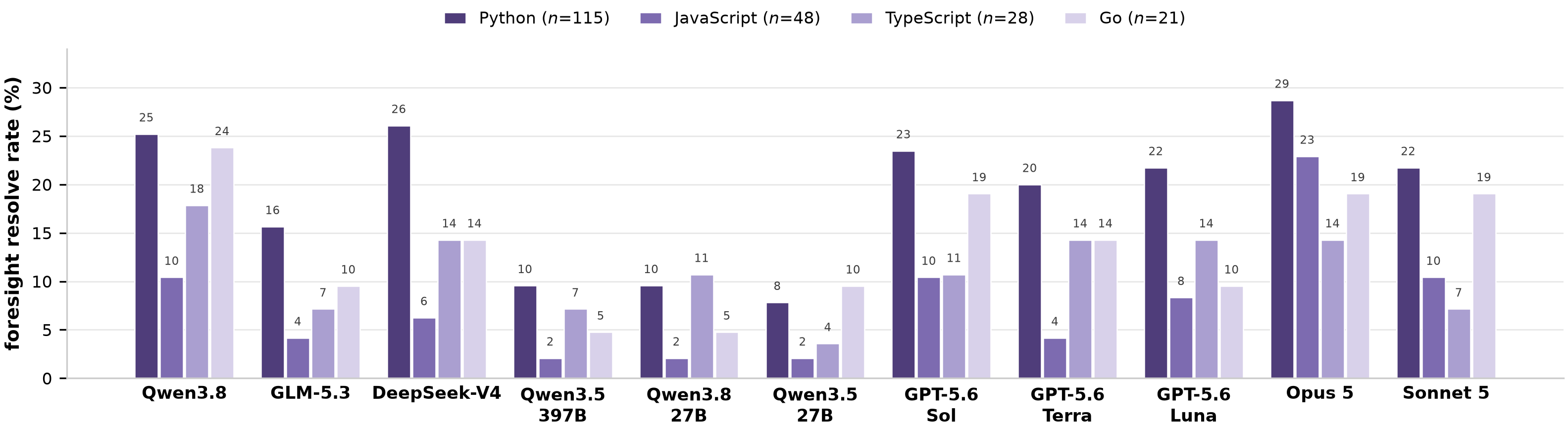}
    \caption{\textbf{Foresight across programming languages.} We report the foresight resolve rate for all eleven models, stratified by repository language. Python is highest for every model and JavaScript lowest or tied-lowest for ten of eleven, with pooled rates of $19.1\%$, $13.4\%$, $11.0\%$ and $7.6\%$ for Python, Go, TypeScript and JavaScript. The Go and TypeScript strata contain $21$ and $28$ instances, so single instances move those bars by $4.8$ and $3.6$ points. JavaScript instances come largely from a single repository.}
    \label{fig:lang_strat_results}
\end{figure}



\end{document}